%% file: paper.tex
\documentclass[acmlarge]{acmart}

\usepackage[utf8]{inputenc}
\usepackage{url}
\usepackage{graphicx}
\usepackage{multirow}
\usepackage{algorithm}
\usepackage[noend]{algpseudocode}

\graphicspath{{img/}{}}

\AtBeginDocument{%
  \providecommand\BibTeX{{%
    \normalfont B\kern-0.5em{\scshape i\kern-0.25em b}\kern-0.8em\TeX}}}

\setcopyright{acmcopyright}
\copyrightyear{2018}
\acmYear{2018}
\acmDOI{10.1145/1122445.1122456}

\acmConference[IMWUT '18]{Woodstock '18: ACM Symposium on Neural
  Gaze Detection}{June 03--05, 2018}{Woodstock, NY}
\acmBooktitle{Woodstock '18: ACM Symposium on Neural Gaze Detection,
  June 03--05, 2018, Woodstock, NY}
\acmPrice{15.00}
\acmISBN{978-1-4503-9999-9/18/06}

\begin{document}

\title{Modulo: Drive-by Sensing at City-scale on the Cheap}

\author{Dhruv Agarwal}
\email{}
\orcid{}
\affiliation{%
  \institution{Ashoka University}
  \streetaddress{}
  \city{}
  \state{}
  \postcode{}
}

\author{Srinivasan Iyengar}
\email{}
\orcid{}
\affiliation{%
  \institution{Microsoft Research India}
  \streetaddress{}
  \city{}
  \state{}
  \postcode{}
}

\author{Manohar Swaminathan}
\email{}
\orcid{}
\affiliation{%
  \institution{Microsoft Research India}
  \streetaddress{}
  \city{}
  \state{}
  \postcode{}
}

\renewcommand{\shortauthors}{Agarwal, et al.}

\begin{abstract}
\input{abstract.tex}
\end{abstract}



\maketitle

\section{Introduction}
\input{introduction.tex}

\section{Background}
\label{sec:background}
\input{background.tex}

\section{Algorithm Design}
\label{sec:algorithm}
\input{algorithm.tex}

\section{Modulo: System Implementation}
\label{sec:implementation}
\input{implementation.tex}

\section{Evaluation Methodology}
\input{methodology.tex}

\section{Experimental Results}

\input{results.tex}

\section{Related Work}
\input{related.tex}

\section{Discussion and Future work}
\input{discussion.tex}

\section{Conclusion}
\input{conclusion.tex}

\bibliographystyle{ACM-Reference-Format}
\bibliography{paper}

\end{document}

%% file: abstract.tex
\textit{Drive-by sensing} is gaining popularity as an inexpensive way to perform fine-grained, city-scale, spatio-temporal monitoring of physical phenomena. Prior work explores several challenges in the design of low-cost sensors, the reliability of these sensors, and their application for specific use-cases like pothole detection and pollution monitoring. However, the process of deployment of a \textit{drive-by sensing} network at a city-scale is still unexplored. Despite the rise of ride-sharing services, there is still no way to optimally select vehicles from a fleet that can accomplish the sensing task by providing enough coverage of the city. In this paper, we propose Modulo -- a system to bootstrap \textit{drive-by sensing} deployment by taking into consideration a variety of aspects such as spatio-temporal coverage, budget constraints. Further, Modulo is well-suited to satisfy unique deployment constraints such as colocations with other sensors (needed for gas and PM sensor calibration), etc. We compare Modulo with two baseline algorithms on real-world taxi and bus datasets. We find that Modulo marginally outperforms the two baselines for datasets with just random-routes vehicles such as taxis. However, it significantly outperforms the baselines when a fleet comprises of both taxis and fixed-route vehicles such as public transport buses. Finally, we present a real-deployment that uses Modulo to select vehicles for an air pollution sensing application. 

%% file: introduction.tex
\textit{Drive-by sensing} is a promising approach to monitoring large geographic areas using a fleet of vehicles equipped to sense relevant quantities. In the recent past, this approach has become quite popular due to improvements in network connectivity and falling sensor costs.  
Moreover, the high proliferation of smartphones, even in the developing world, allows leveraging in-built sensors on mobile phones of willing participants to sense physical phenomena. Particularly in an urban setting, \textit{drive-by sensing} has led to several interesting works that can empower individuals and other stakeholders (i.e., municipal authorities) to make informed decisions. Examples of such prior work include, but are not limited to, monitoring traffic congestion~\cite{mohan2008nericell}, detecting potholes~\cite{eriksson2008pothole}, sensing air quality~\cite{HSWT2011a,gao2016mosaic}, recognizing unsafe pedestrian movement~\cite{datta2014towards}, recording parking violations~\cite{he2018detecting} and identifying available parking spaces~\cite{mathur2010parknet,rong2018parking}, etc.

These applications have proved their usability in real-world deployments and have consequently progressed the concept of \textit{drive-by sensing}. It is also the case that larger the scale of deployment of such systems, the more profound knowledge one can get about the sensed entity. To that effect, city-scale deployment of such solutions that consolidate the data from multiple geographically separated sensor nodes 
is desirable for urban planning. For example, a mobile phone of a single commuter driving along a road can flag vehicular congestion. However, such data accumulated from commuters in major road segments of a city permit the use of analytics to route traffic effectively. Likewise, a few low-cost PM2.5 sensors can report air pollution levels around their locations. But a dense network of these sensors spread all over a region can help identify hotspots. Resulting fine-grained data on pollution levels enables running causality studies and helps design corrective steps to improve air quality. Smart city initiatives around the world are banking on getting actionable insights through similar widespread participatory sensing~\cite{articlepart}.

\textit{Drive-by sensing} at city-scale offers several additional benefits as compared to traditional static sensing. First, unlike static sensors which are installed to measure the entity of interest from individual locations, \textit{drive-by sensing} can be employed to cover a wider area. Thus, at a fraction of a cost, one can get \textit{coverage} equivalent to multiple static monitors. In turn, \textit{drive-by sensing} reduces the administrative burden of managing a more extensive network. Moreover, some sensors require continuous on-field calibration to correct for any sensor drifts. With \textit{drive-by sensing}, vehicles can potentially rendezvous across a static reference-grade monitor a few times each day. Thus, the process of calibrating low-cost mobile sensors does not necessitate the suspension of their operation. On the contrary, given the necessity of regular calibration, static sensors would have numerous periods of downtime over their lifespan for calibration through manual colocation with expensive monitors. Additionally, in \textit{drive-by sensing}, a recently calibrated mobile sensor can calibrate other sensors when they transit each other. A few papers have utilized these \textit{multi-hop calibrations} for gas~\cite{maag2017scan} and particulate matter~\cite{fu2017multihop} sensors. 

\textit{Drive-by sensing} inherently requires a fleet of vehicles that regularly move around the geographical area of interest. Several papers have used public transportation as their choice of the vehicle fleet~\cite{HSWT2011a}. With predictability in routes~\footnote{Public transport vehicles operate on fixed routes. These routes and their respective schedules are made available to the residents of the city to help them plan their commute.} and lower cost of operation, one can summon this resource for monitoring urban locations. But, many cities around the world do not have a good public transport system and therefore, cannot guarantee sufficient coverage~\cite{radio}. Fortunately, over the past few years, ride-sharing services~\footnote{Companies such as Uber, Lyft, Ola, Didi} have gained tremendous popularity and their vehicle fleets have emerged as a credible alternative for \textit{drive-by sensing}. For example, as the largest of such ride-sharing companies, Uber is operating in over 600 cities spread over 65 countries and is servicing around 15 million trips per day~\cite{cnet}. These numbers are expected to grow in the near future. 

Although seemingly straightforward, piggybacking on either or both of public transport and ride-sharing services as the medium for city-scale sensing raises several complications. 
First, the frequency of spatio-temporal data collection depends on the type of sensing application.
Further, the frequency itself may not be uniform as the sensing application might need more data during a specific time of the day or at a specific location. Certain applications (SO$_{2}$ or O$_{3}$ sensing) need deployment that accomplishes regular colocations with reference-grade monitors for calibration. We will discuss these complicating factors in more detail in Section~\ref{sec:background}. 

In this paper, we account for these factors and select the optimal set of vehicles that can cover a wide area. Our approach can bootstrap the process of sensor deployment in any city by examining its historical vehicular mobility patterns while simultaneously satisfying the pre-decided collection constraints. Specifically, we make the following key contributions:
\begin{itemize}
   \item \textbf{Vehicle Selection Algorithm:} We formulate the problem of selecting the optimal set of vehicles as an integer linear program. Further, we introduce several relevant extensions suitable for different \textit{drive-by sensing} applications by changing either or both the objective and the constraints of this linear program. As the problem is \textbf{NP-Hard}, we present a greedy algorithm that gives the best-possible polynomial-time approximation algorithm for our problem. 
    \item \textbf{System Implementation:} We introduce Modulo, our system for selecting the optimal set of vehicles for any \textit{drive-by sensing} application. Modulo provides in-build support for using our greedy algorithm while considering various deployment-level details. Notably, for sensing applications that require calibration (gas and particulate matter sensors), Modulo uses geohash --- a geospatial indexing approach --- to quickly find mobile colocations with other sensors. Modulo is released as an open-source  Python library for the community to utilize our approach for individual \textit{drive-by sensing} applications. 
    \item \textbf{Detailed evaluation:} We benchmark Modulo against two baseline approaches for vehicle selection. We evaluate the performance on real-world datasets from San Francisco and Rome. Moreover, we use two kinds of datasets: one of taxis that move randomly across the city, and the other one containing buses from the public transportation system that follow fixed routes over a long period of time. We also conducted a case-study involving air pollution monitoring in a southern Indian city using the vehicles selected by Modulo.
\end{itemize}


%% file: background.tex
In this section, we will provide a brief background on the unique distinguishing characteristics of \textit{drive-by sensing} to monitor urban environments. We also elaborate on how the spatiotemporal variability observed in the measured entity (such as pollution, potholes, traffic congestion, etc.) dictate the strategy utilized for sensing them. Finally, we formulate the problem of \textit{drive-by sensing} deployment, which takes into consideration the above factors. 

\subsection{Drive-by sensing}
Public transport vehicles (i.e., buses, trams) or other vehicles driven by working individuals with a set workplace and schedule have \textbf{fixed routes}~\footnote{Public transport buses may not be assigned the same route each day. Here, we assume that the sensor node can be transferred to the buses that are plying on the selected (fixed) routes}. With these vehicles, there is almost complete certainty on the coverage achieved based on historical data. In cities with efficient public transport system, using just the \textbf{fixed route} vehicles might suffice as the sensors placed on them can cover major portions of a city.

Whereas in some cases, we might have to rely on cabs operated by ride-sharing platforms. The movement of these vehicles may vary significantly from one day to another and have \textbf{random routes}. Thus, this stochasticity in motion may lead to non-uniform coverage in the different parts of the city. However, with the proper selection of vehicles that cover more distance on varied routes, one can reduce the probability of sparse sensing. A composite approach would involve deploying sensors on both fleets of vehicles. Naturally, the choice of selecting the vehicles (having either fixed or random routes) will vary from city to city.  

\begin{table}[t]
\begin{tabular}{|c|c|c|c|c|}
\hline
\multirow{2}{*}{\textbf{\begin{tabular}[c]{@{}c@{}}Drive-by sensing \\ application\end{tabular}}} & \multicolumn{2}{c|}{\textbf{Granularity}} & \multicolumn{2}{c|}{\textbf{Variability}} \\ \cline{2-5} 
 & \textbf{Spatial} & \textbf{Temporal} & \textbf{Spatial} & \textbf{Temporal} \\ \hline
Ozone gas & Low & Medium & Uniform & Uniform \\ \hline
Parking spot availability & High & High & \begin{tabular}[c]{@{}c@{}}Weighted towards city-\\ centers and public areas\end{tabular} & \begin{tabular}[c]{@{}c@{}}Weighted higher \\ during peak times\end{tabular} \\ \hline
Particulate matter & Medium & Medium & Uniform & Uniform \\ \hline
Pothole detection & High & Low & Uniform & \begin{tabular}[c]{@{}c@{}}Weighted; bursts in \\ data collection okay \end{tabular}\\ \hline
Traffic Congestion & Medium & Medium & Uniform & \begin{tabular}[c]{@{}c@{}}Weighted higher\\ during peak hours\end{tabular} \\ \hline
\end{tabular}
\caption{Drive-by sensing application needs}
\label{table:granularity}
\end{table}

\subsection{Spatiotemporal granularity and variability in the sensed entity}
Here, we define two complementary properties that underlay any \textit{drive-by sensing} application --- \textit{granularity} and \textit{variability}. These properties apply to both the spatial and temporal dimensions. Understanding these properties will lead to a consequent data collection that is representative of the characteristics exhibited by the specific sensing use-case. Below, we define these two properties and explain how they are relevant in deciding on a deployment strategy. 

\begin{itemize}
    \item \textbf{Granularity:} Granularity is the expected frequency of data collection required by a sensing application. As discussed, granularity can be defined separately for space and time dimensions. Requiring high temporal granularity would translate to sensing every few seconds or minutes. Sensing at a low temporal granularity equates to recording data every few days. Similarly, high spatial granularity needs recording data every few meters. Whereas, low spatial granularity means collecting data every few kilometers.
    \item \textbf{Variability:} Variability is the extent to which changes in the granularity of the data collection that is can be tolerated (or needed) for a sensing application. Again, variability can be defined separately for space and time dimensions. Requiring uniform spatial variability suggests having similar granularity for different regions in the sensed data. Correspondingly, weighted spatial variability means that the data collection changes with different areas.
\end{itemize}


The spatial and temporal granularity needed to perform \textit{drive-by sensing} vary depending on the type of applications. For example, ozone gas concentrations do not change over a few kilometers. Thus, building \textit{drive-by sensing} network with low spatial granularity will suffice.  However, detecting pothole on roads might need multiple transits from various vehicles, preferably moving on separate lanes to tag the road segments correctly. At the same time, the potholes may not change drastically over a few hours or even days. Thus, for detecting potholes, we can do with lower temporal granularity. Interestingly, some applications need both higher spatial and temporal granularity. An example application would be identifying available parking spots using cameras mounted on cabs. Nonetheless, such an application can have variability in sensing. For example, the higher granularity is necessary only during the peak hours and at city-centers where there is a contention on the available slots. Thus, the sensing solution needs to be weighted more towards regions and times that tend to be more crowded. Whereas, in the case of particulate matter, we would like to have uniformity in spatial and temporal variability. Table~\ref{table:granularity} lists several applications and their associated \textit{drive-by sensing} needs in terms of granularity and variability.


%% file: algorithm.tex
We start with formulating the problem of efficient city-scale deployment of low-cost sensors on the fleet of vehicles. Given a sensing application, one must decide on the granularity and variability requirements. Based on these requirements, one can leverage vehicles willing to participate in the \textit{drive-by sensing} exercise. Further, the optimal number of vehicles with either or both fixed and random routes can be considered using their mobility patterns. These patterns include schedules (or past transit times) of public transport vehicles, the historical transit data of vehicles registered with the ride-sharing services, etc. 

Let $D$ represent the set of segments that are obtained by spatially partitioning the city under consideration for \textit{drive-by sensing}. The segments could be line slices covering the entire road network or polygonal regions covering the whole city. Further, the spatial granularity specified decides to the cardinality of the set of segments ($|D|$). Higher the granularity, higher is the value of $|D|$. Similarly, let $T$ represent the set of time intervals that are obtained by temporally partitioning the historical vehicular mobility data. Again, the number of the time intervals ($|T|$) depends on the minimal temporal granularity specified. 

Now, let us consider, $n$ vehicles with known mobility pattern represented as a collection of sets $V = \{V_{1}, V_{2},....,V_{n}\}$. A set in this collection, represented by $V_{i}$ where $i \in [1,..,n]$, contains the multiple tuples - $\langle d,t \rangle$, which represents the segment index $d \in D$ and time interval index $t \in T$. A presence of a tuple indicates that the vehicle visited segment $d$ at time interval $t$. Our objective is to increase the coverage of the sensing setup containing the selected vehicles $V^{\prime} \subseteq V$, such that $|V^{\prime}| \leq m$. Here, $m$ is the maximum number of vehicles that can be summoned based on a predefined budget and $m \leq n$. Mathematically, we want to maximize -  

\begin{equation}
\left\vert \bigcup_{V_{i} \in V^{\prime}} V_{i} \right\vert
\end{equation}

Essentially, we are maximizing the union of the set containing $\langle d,t \rangle$ tuples visited by the selected vehicles. Obviously, in this formulation, we do not get any benefit by visiting a tuple more than once. 

\subsection{Integer Linear Programming Formulation}
Here, we present a linear programming formulation that maximizes the term described earlier. 
Let $x_{i}$ be a binary decision variable representing if the $i \in [1,...,n]$ vehicle is selected by our optimization. 
Further, let $m$ be the maximum budget of vehicles that can be selected for a given \textit{drive-by sensing exercise}. Let $O_{d,t}$ represent the occupancy obtained from these selected vehicles at time interval $t \in T$, i.e., the set of all time intervals, and $d \in D$, the set of segments. This variable again takes binary values. $O_{d,t}=1$, if the segment $d$ at time interval $t$ was occupied by any of the selected vehicle. Let $C_{i,d,t}$ represent a known binary parameter that describes if the $i^{th}$ vehicle travelled to segment $d$ at time interval $t$. Based on these parameters and variables, we define our integer linear programming formulation as: 

\begin{align*}
\max  &\quad  \sum_{ d\in D,t\in T} O_{d,t} & \quad \text{(Objective function)} \\
\text{subject to:}&\quad \sum_{i \in [1,...,n]} x_{i} \leq m &  \quad \text{(Budget constraint)},\\
& \quad \sum_{i \in [1,...,n]} (C_{i,d,t} \cdot x_{i}) \geq O_{d,t} \quad \forall d\in D, t\in T & \quad \text{(Coverage-Occupancy constraint)}, \\
&\quad O_{d,t} \in \{0,1\} \quad \forall d\in D, t \in T & \quad \text{(Occupancy binary constraint)},\\
&\quad x_{i} \in \{0,1\} \quad \forall i \in [1,...,n]  & \quad \text{(Vehicle binary constraint)}\\
\end{align*} 

We expand on the choice of the objective function and the constraints mentioned above as follows:- 

\begin{itemize}
    \item \textbf{Objective function:} We want the selected vehicles to travel to as many segments at different time intervals to ensure maximum coverage. Thus, we define our objective function to maximize the sum of the binary variable occupancy  $O_{d,t}$ over every $d \in D$ and $t \in T$. 
    \item \textbf{Budget constraint:} Out of all the vehicles for which the mobility patterns are known, we want to select a subset of them based on a predefined budget. Hence, the sum over the binary variable $x_{i}$, representing selected vehicles, cannot cross the budget allocated ($m$). 
    \item \textbf{Coverage-Occupancy constraint:} If the occupancy variable $O_{d,t}>0$, then at least one of the selected vehicles was present in segment $d$ at time interval $t$. This constraint ensures that multiple transits from the selected vehicles over a specific segment $d$ and time interval $t$ are not counted multiple times. 
    \item \textbf{Occupancy binary constraint:} As defined earlier, occupancy $O_{d,t}$ is a binary variable. This constraint ensures that the variable can only take one of the two values of 0 and 1. 
    \item \textbf{Vehicle binary constraint:} As defined earlier, the variable $x_{i}$ is a binary variable. This constraint ensures that the decision variable on the selection of vehicles can only take one of the two values of 0 and 1. 
\end{itemize}

\subsubsection{Algorithm Analysis}
The above formulation exactly matches the classical \textit{maximum coverage problem} --- a widely studied problem in the theoretical computer science and operations research community. Unfortunately, the \textit{maximum coverage problem} is \textbf{NP-hard}~\cite{hochbaum1997approximating}. Thus, our problem cannot be solved exactly in polynomial time, if \textbf{P $\neq$ NP}.

\subsubsection{Calibration considerations} 
The described integer programming formulation provides a general framework for selecting vehicles for \textit{drive-by sensing}. However, a few sensing applications require placing additional constraints. For example, in the case of city-scale deployment of low-cost gas or particulate matter sensors, one needs to ensure that they are calibrated on-field regularly. This obligation necessitates the low-cost sensors to rendezvous around reference-grade pollution measuring stations~\footnote{These stations use gravimetric methods to measure pollution. Example: Beta Attenuation Monitors.} or other low-cost sensors that have been recently calibrated. To guarantee calibration, we can easily extend the above formulation by ensuring that the selected sensors have mobile colocations with either or both of reference-grade and other low-cost sensors. Let us consider the parameters $Lb_{i}$ and $Ls_{i}$ represent the number of colocations of the $i^{th}$ vehicle with reference-grade and other low-cost sensors respectively. Then we can add the following constraints - 

\begin{align*}
& \quad Lb_{i} \cdot x_{i} \geq b \quad \forall i\in [1,...,n] & \quad \text{(Reference colocations constraint)}, \\
& \quad Ls_{i} \cdot x_{i} \geq s \quad \forall i\in [1,...,n] & \quad \text{(Low-cost sensor colocations constraint)}, \\
\end{align*} 

Above, $b$ and $s$ represent the minimum number of colocations needed for selected vehicles with reference-grade and other low-cost sensors, respectively.

\subsubsection{Budget Minimization Formulation}
We can easily change the above formulation from maximizing coverage for a given budget to minimizing the budget necessary to meet a specific amount of coverage. For this modification we can replace the objective functions and the budget constraint, while keeping the other constraints as is in the following way: 
\begin{align*}
\min  &\quad  \sum_{i \in [1,...,n]} x_{i} & \quad \text{(Budget minimizing objective function)} \\
\text{subject to:}&\quad \sum_{ d\in D,t\in T} O_{d,t} \geq k &  \quad \text{(Minimum coverage constraint)}
\end{align*} 

\subsubsection{Weighted Coverage Formulation}
As described in section~\ref{sec:background}, some sensing applications might require the variability in coverage to be weighted at certain spatial and temporal levels (see the "Parking Spot Availability" sensing application in Table~\ref{table:granularity}). The above integer linear programming formulation can easily accommodate this by modifying the objective function in the following way:

\begin{align*}
\max  &\quad  \sum_{ d\in D,t\in T} O_{d,t}\cdot W_{d,t} & \quad \text{(Weighted objective function)} 
\end{align*}

Here, $W_{d,t}$ represents the weights given to the segment $d \in D$ at time interval $t \in T$. For uniform variability, the $W_{d,t} =1$  $\forall d\in D$ and $t \in T$. The weighted maximum coverage problem is also widely studied in the literature~\cite{nemhauser1978analysis}.

\subsubsection{Variable cost of deployment}
In some cases, the deployment cost associated with each type of vehicle can be variable. For example, the city authorities might have different incentive mechanisms for individual car owners participating in \textit{drive-by sensing} compared to drivers of the ride-sharing platforms. Again, the above formulation can handle the difference in incentive mechanism by changing the budget constraint as follows: 

\begin{align*}
&\quad \sum_{i \in [1,...,n]} B_{i} \cdot x_{i} \leq p &  \quad \text{(Budget constraint)},\\
\end{align*}  

Here, the parameter $B_{i}$ represents the cost of deployment that covers the incentive mechanisms for the $i^{th}$ vehicle. Whereas, the parameter $p$ serves as the overall expenditure limit for given \textit{drive-by sensing} application. The discussion on the budgeted case is beyond the scope of this paper. Interested readers are requested to read the work by Khuller et al.~\cite{khuller1999budgeted} for a more details. 

\subsection{Greedy Approximation}
\label{sec:greedy_approximation}
As discussed earlier, the maximum coverage problem is \textbf{NP-hard}. Fortunately, there exists a greedy heuristic that provides a solution within an approximation ratio of $1-\frac {1}{e}$, i.e., the best polynomial-time approximation algorithm --- unless \textbf{P = NP} --- for both unweighted and weighted maximum coverage problem~\cite{hochbaum1997approximating,nemhauser1978analysis}.

We modify the same greedy algorithm (see Algorithm~\ref{alg:greedy}) to include additional constraints such as to ensure a minimum number of reference and low-cost sensor colocations. This algorithm computes a list called $vehicles$ that maximizes the weighted coverage. Similar to the notations used earlier, we have a predefined budget $m$, a binary parameter $C_{i,d,t}$, which indicates if the $i^{th}$ vehicle was in segment $d$ at time interval $t$. $O_{d,t}$ is a binary variable that indicates if the segment $d$ at time interval $t$ was covered by the selected vehicles. Functions  $referenceColocations(j)$ and $sensorColocations(j)$ return the number of colocations of the $j^{th}$ vehicle with reference-grade monitors and other low-cost sensors, respectively.

\begin{algorithm}
\caption{Greedy Sensor Deployment for Drive-by Sensing}\label{alg:greedy}
\begin{algorithmic}[1]
\State \textbf{Parameters:} i) $C_{i,d,t} \forall i \in [1,...,n]$, $ d \in D$, $t \in T; \quad$ ii) $minRefColocations; \quad$ iii) $minSenColocations \quad$ iv) $m$
\State \textbf{Initialize:} i) $ currentOccupancySum = 0;  \quad$ ii) $vehicles = []; \quad$ iii) $O_{d,t} = 0  \forall d \in D \quad t \in T$
\For{i in [1,...,m]} 
    \For{j in [1,...,n]}
        \If{$referenceColocations(j)>=minRefColocations$}  \textit{\#Optional Constraint}
        \If{$sensorColocations(j)>=minSenColocations$} \textit{\#Optional Constraint}
        \State $\dot O_{d,t} = O_{d,t} \cup C_{j,d,t} \quad \forall d \in D \quad t \in T$
        \If{$currentOccupancySum < \sum_{d \in D, t\in T} \dot O_{d,t} \cdot W_{d,t} \quad $}
            \State $selectVehicleIndex = j$
            \State $currentOccupancySum = \sum_{d \in D, t\in T} \dot O_{d,t}  \cdot W_{d,t}$
        \EndIf
        \EndIf
        \EndIf
   \EndFor
   \State $O_{d,t} = O_{d,t} \cup C_{selectVehiclesIndex,d,t} \quad \forall d \in D \quad t \in T$ 
   \State $vehicles.append(selectVehiclesIndex)$
\EndFor
\end{algorithmic}
\end{algorithm}

\subsubsection{Greedy formulation for budget minimization}
Rather than coverage maximization, one can get the set of vehicles that minimize the budget for a given minimum expected coverage. For this, we can change line 3 in the Algorithm~\ref{alg:greedy} to ``$\textbf{while} \sum_{d \in D, t \in T} O_{d,t} \cdot W_{d,t} < k \quad \textbf{do}$'', where $k$ is the minimum coverage expected. 

\subsubsection{Handling dynamism in deployments}
In any \textit{drive-by sensing} exercise, there might be cases where one would have to swap the sensors from one vehicle to the other for several reasons. For example, there could be churn in the driver pool of a ride-sharing service or a few participating cab drivers may not want to continue with the sensing application. Further, with an increased budget, there is scope for adding new vehicles to the sensing fleet. For long-term deployments of \textit{drive-by sensing}, such dynamism would be commonplace. Thus, it is essential to support cases where one would like to have a new list of vehicles for an incremental deployment.  

For this, we can modify the existing greedy algorithm by changing the  parameters and initializations to reflect the incremental case.  In parameters, $C_{i,d,t}$ will be a binary parameter showing remaining vehicles that are not part of the sensing application, and $m$ will be the remaining or additional budget available. Likewise, in initializations, $vehicles$ will contain the current set of vehicles and $O_{d,t}$ will have the occupancy from these vehicles. Whereas, $currentOccupancySum$ is set to $\sum_{d \in D, t\in T} O_{d,t} \cdot W_{d,t}$, calculated using the current vehicles.

%% file: implementation.tex

\begin{figure}
    \centering
    \includegraphics[width=\textwidth]{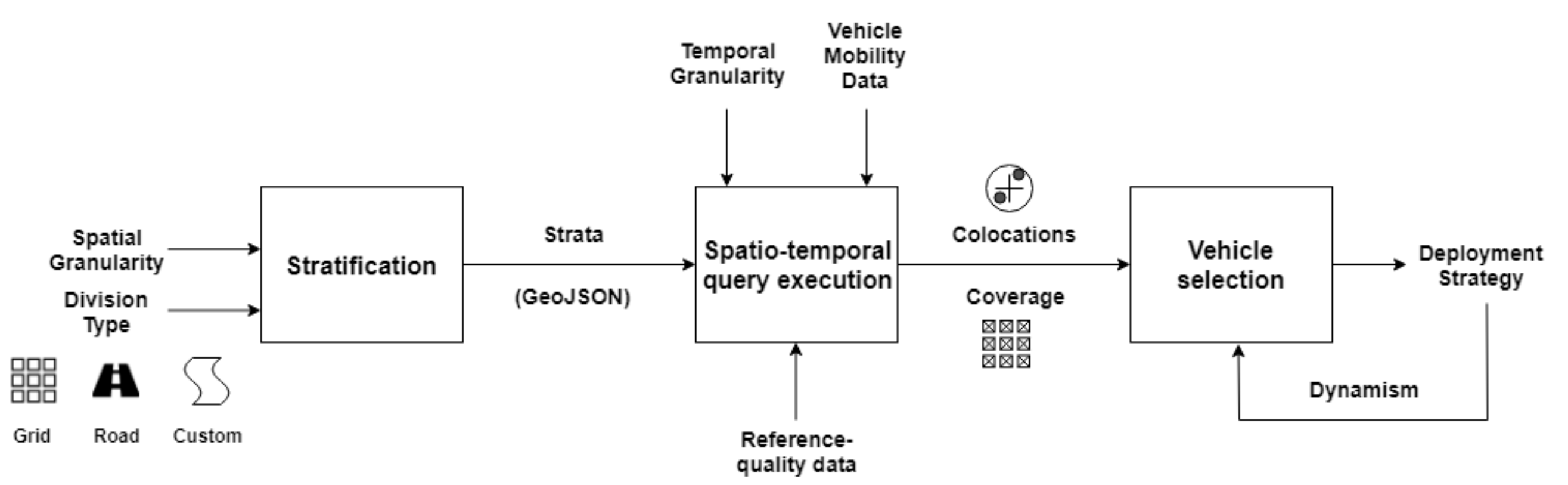}
    \caption{Pipeline of Modulo}
    \label{fig:pipeline_modulo}
\end{figure}

In this section, we describe Modulo, a system to identify the optimal set of vehicles needed by operators of large-scale \textit{drive-by sensing} networks (from hereon referred just as operators). Modulo is designed to be \textit{application-agnostic}, i.e., it is general enough to serve a plethora of sensing sensing use-cases.  Figure \ref{fig:pipeline_modulo} shows the complete pipeline of Modulo consisting of three key steps:  

\begin{enumerate}
\item \textbf{Stratification} creates strata over the geographical region based on the appropriate division type entered by the operator. 
\item \textbf{Spatio-temporal Query Execution} primarily takes historical mobility data entered for different vehicles types as input. It computes the coverage of each vehicle over a specified spatial granularity and temporal granularity using the strata created in the previous step. 
\item \textbf{Vehicle Selection} uses the data computed for coverage along with the user-defined parameters (such as the maximum number of vehicles, etc.) to select the optimal set of vehicles using the greedy algorithm described in section \ref{sec:greedy_approximation}.
\end{enumerate}
Below, we describe each step in detail.



\subsection{Step 1 - Stratification}

In this step, the operator inputs a list of coordinates (latitude and longitude) encompassing the chosen geographical area to partition into smaller regions. Further, the operator provides the spatial granularity required by the sensing application. Section~\ref{sec:background} presents brief guidelines for the granularity required by various \textit{drive-by sensing} use-cases. Additionally, the operator can choose one of the two types of strata --- i) Square-shaped grids, or ii) Road segments. This feature is provided as different sensing use-cases may require different kinds of stratification of the city. For air pollution sensing, it could be uniformly sized grids. Whereas, for pothole detection, it could be segments of the road network. 

Modulo has in-built support to partition the geographic area into uniform grids and roads using the spatial granularity provided. Modulo accomplishes partitioning into square-shaped grids using the Turf.js JavaScript library~\cite{turfjs}. To identify roads in a geographical area,  Modulo uses the Mapbox Static Maps API~\cite{mapbox}. Essentially, it retrieves 1280 px $\times$ 1280 px sized static images of all the grids overlayed on a map of the road network of the city. Modulo applies basic image processing to ascertain the percentage of each grid that is covered with roads and considers the ones that are above a certain percentage threshold. Further, segmentation of roads is also performed using Mapbox APIs. The resulting strata are outputted as a GeoJSON \cite{gillies2016geojson} file. Alternatively, the operator can provide a custom  GeoJSON file with partitioning of the geographical region as GeoJSON encoded polygons. Modulo assigns each polygon representing the various types of strata a unique stratum ID, which gets embedded in the exported GeoJSON file.

\subsection{Step 2 - Spatio-temporal Query Execution}
In this step, the operator inputs historical mobility data of the fleet that they want to use for deployment. The operator also input a temporal granularity that is required for their use case. The spatial and temporal granularity together decide the coverage of the deployment.
Optionally, the administrator can also input the locations and time frequencies of any reference-grade sensors they may have deployed across the city. This is important for use-cases like air pollution sensing, where the low-cost mobile sensors need to be tracking calibrated by multi-hop colocations from the reference-grade sensors.

Modulo handles the historical data provided in three stages:
\begin{itemize}
    \item The raw data is parsed, and for each record, the timestamp, GPS location, and vehicle ID are inserted into a NoSQL MongoDB \cite{mongodb} database.
    \item A spatio-temporal compound index is created on this database for efficient querying of data. MongoDB implements a geospatial index by calculating a geohash on the locations of the records. A geohash divides the earth's surface into grids and encodes a location into an alphanumeric string. This string has the nice property that places near to each other will have similar prefixes~\cite{geohash}.
    \item All the records are assigned a \texttt{stratum\_id} depending on the stratum that they fall under. This can be done efficiently because of the geospatial index and the GeoJSON resulting from step 1 of the pipeline. 
    \item All the records are also assigned a \texttt{time\_interval\_id} depending on the temporal granularity. 
\end{itemize}

After these three stages, every NoSQL record in the database has the following structure:
\begin{verbatim}
{
    vehicle_id: <int>,
    stratum_id: <int>,
    time_interval_id: <int|Unix timestamp>,
    timestamp: <int|Unix timestamp>,
    location: {
        "type": "Point",
        "coordinates": [<float|longitude>, <float|latitude>]
    }
}
\end{verbatim}

With this setup in place, Modulo runs queries to find the coverage achieved by each vehicle in the dataset as per their \texttt{stratum\_id}'s and \texttt{time\_interval\_id}'s. 
Further, if needed, Modulo also runs queries to find colocations in the mobility data. These colocations are separated into two types: with reference-grade sensors and with other vehicles in the mobility dataset. The data thus generated is fed into the next stage in the pipeline to improve upon the achieved coverage.

\subsection{Step 3 - Vehicle Selection}
In the final step of the pipeline, Modulo runs a greedy approximation as detailed in Section~\ref{sec:greedy_approximation}. The result of this algorithm is a deployment strategy for the network. More specifically, the greedy algorithm returns a \textbf{set of vehicles} that should be used for deployment of the drive-by sensing network for improved coverage and colocation count. This final step of the pipeline can be run every few days by the operator in order to account for the dynamism in the sensor network deployment.

\subsection{Open-source Python Library}
We release Modulo as an open-source Python Library~\footnote{We will provide the link to the public GitHub repository for Modulo here in the camera-ready submission.}. This library exposes APIs in the form of three ready-to-use functions that map to the different steps in the Modulo pipeline. These functions are explained briefly in table \ref{tab:modulo_python_library}. We have explained the specifics of input formats (such as JSON specification for vehicle mobility data, etc.) in the documentation accompanying the library in the source code. Further, the library repository also contains easy-to-follow examples in the form of iPython notebooks.

\begin{table}[h]
\begin{tabular}{|c|l|l|}
\hline
\textbf{Function Name} & \multicolumn{1}{c|}{\textbf{Input}} & \multicolumn{1}{c|}{\textbf{Output}} \\ \hline
\texttt{stratify} & \begin{tabular}[c]{@{}l@{}}\texttt{spatial\_granularity}: in meters\\\hspace{1cm} \textbf{type}: int, required\\\hspace{1cm} \textbf{default}: 100\\  \texttt{division\_type}: "grid", "road", "custom"\\\hspace{1cm} \textbf{type}: str, required\\\hspace{1cm} \textbf{default}: "grid"\\ \texttt{geojson}: path to file\\\hspace{1cm} \textbf{type}: str, conditionally required\\\hspace{1cm} \textbf{default}: None\end{tabular} & \begin{tabular}[c]{@{}l@{}}\textbf{Type}: GeoJSON\\ \textbf{Content}: strata boundaries\\ embedded with stratum IDs\end{tabular} \\ \hline
\texttt{execute\_queries} & \begin{tabular}[c]{@{}l@{}}\texttt{temporal\_granularity}: in seconds\\\hspace{1cm} \textbf{type}: int, required\\\hspace{1cm} \textbf{default}: 3600 (i.e. 1 hour)\\ \texttt{mobility\_data}: path to JSON file\\\hspace{1cm} \textbf{type}: str, required\\\hspace{1cm} \textbf{default}: None\\ \texttt{strata\_data}: path to GeoJSON file\\\hspace{1cm} \textbf{type}: str, required\\\hspace{1cm} \textbf{default}: None\end{tabular} & \begin{tabular}[c]{@{}l@{}}\textbf{Type}: JSON\\ \textbf{Content}: 1) coverage\\ data, 2) colocation data\end{tabular} \\ \hline
\multicolumn{1}{|l|}{\texttt{select\_vehicles}} & \begin{tabular}[c]{@{}l@{}}\texttt{coverage}: path to coverage file\\\hspace{1cm} \textbf{type}: str, required\\\hspace{1cm} \textbf{default}: None\\  \texttt{colocations}: path to colocations file\\\hspace{1cm} \textbf{type}: str, optional\\\hspace{1cm} \textbf{default}: None\end{tabular} & \begin{tabular}[c]{@{}l@{}}\textbf{Type}: list\\ \textbf{Content}: vehicle IDs to\\ be selected\end{tabular} \\ \hline
\end{tabular}
\caption{Functions exposed as part API in the Modulo open-source library}
\label{tab:modulo_python_library}
\end{table}





\begin{figure*}[t]
    \centering
    \begin{tabular}{ccc}
    \includegraphics[height=1.9in]{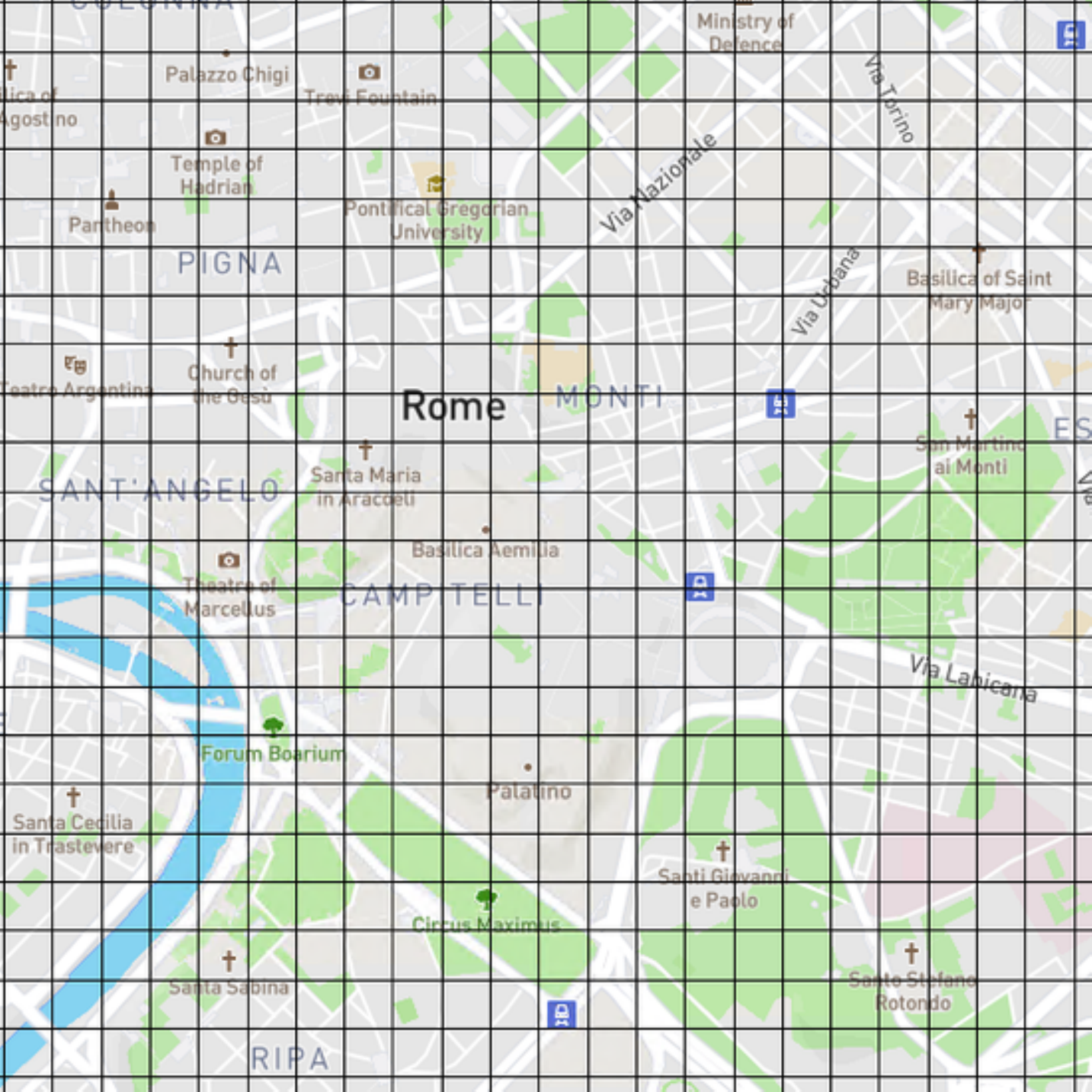} &
    \includegraphics[height=1.9in]{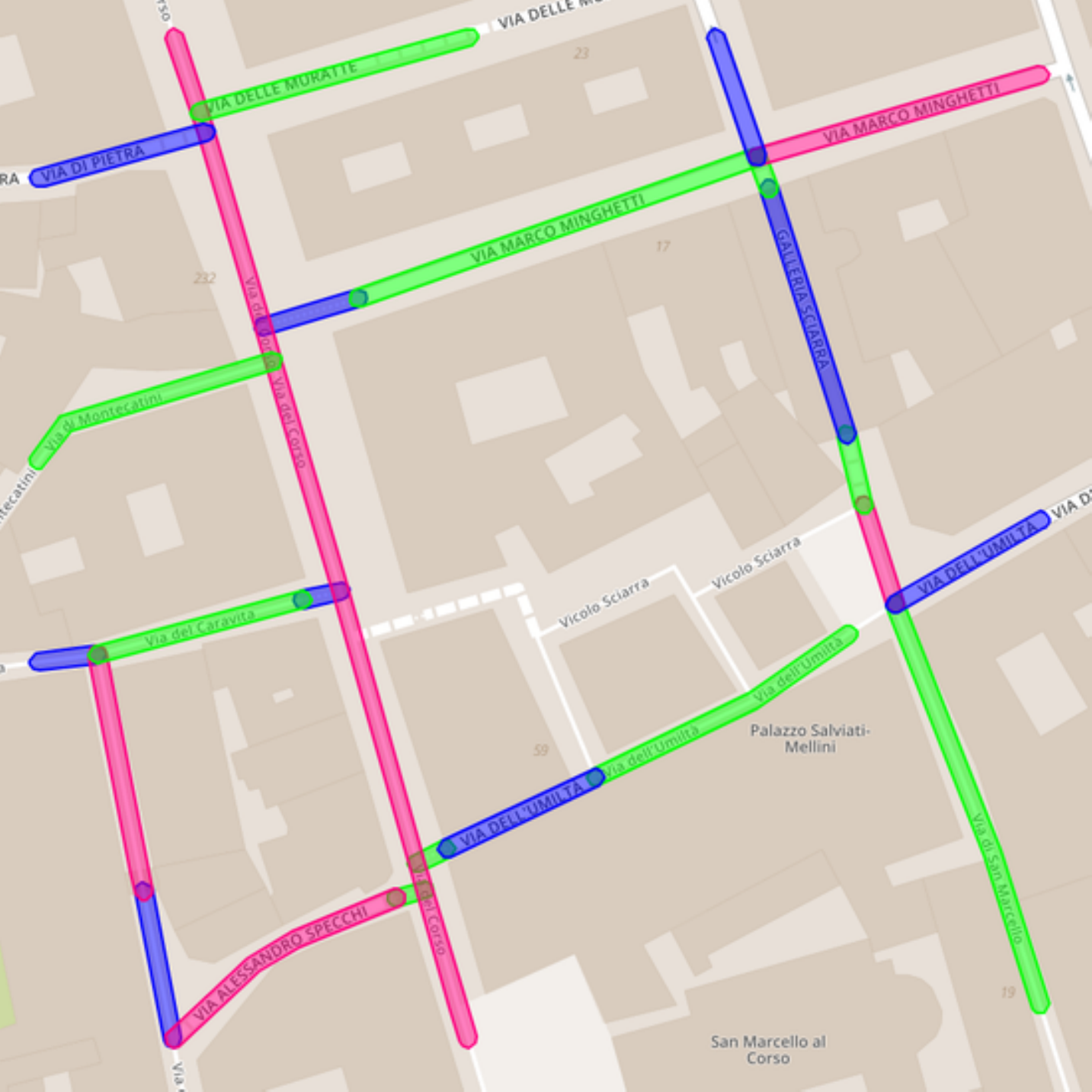} &
    \includegraphics[height=1.9in]{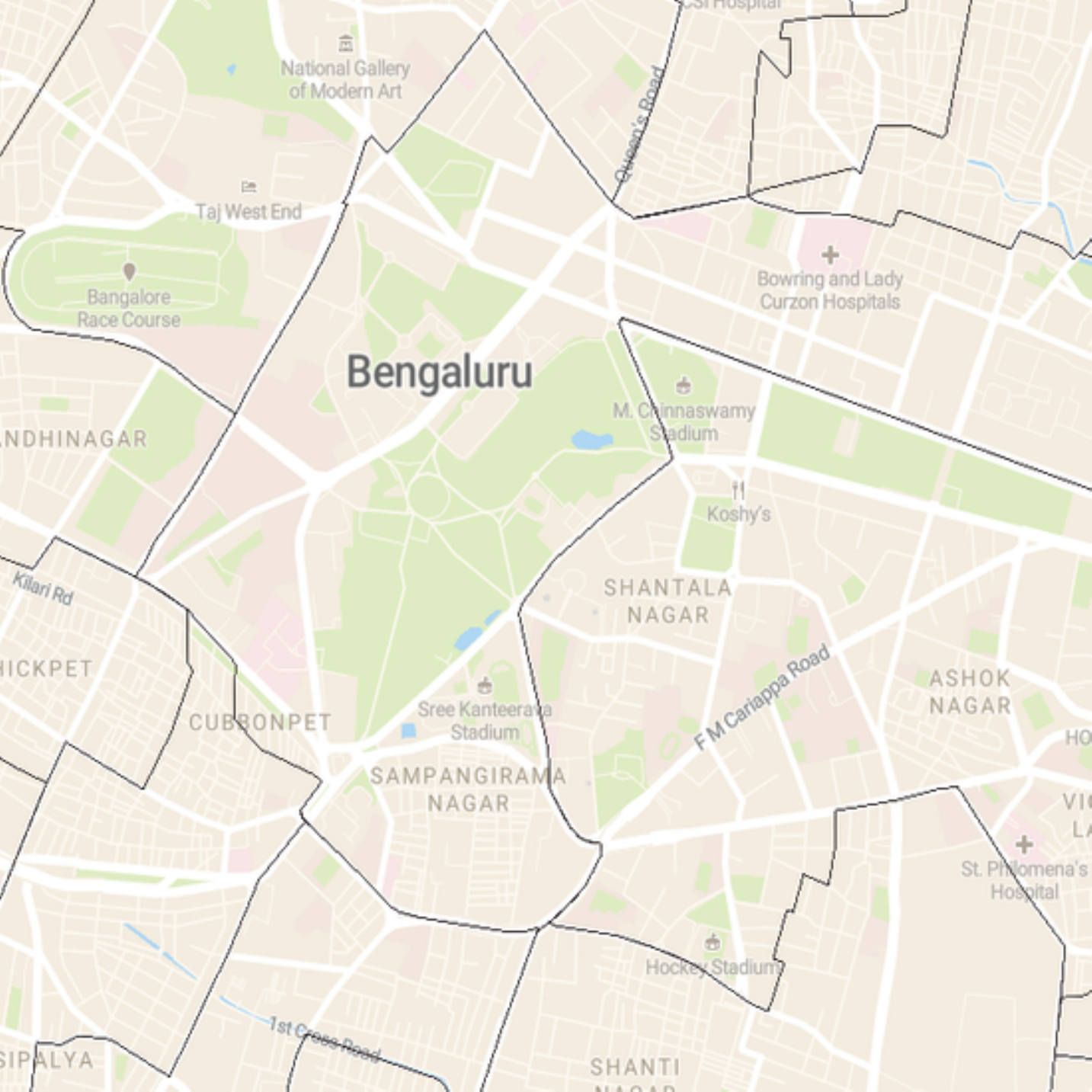} \\
    (a) Grid division in Rome & (b) Road segmentation in Rome & (c) Wards in Bangalore \\
    \end{tabular}
    \caption{This figure shows the different division types enabled by Modulo. (c) shows the division of Bangalore on the basis of administrative boundaries defined in a GeoJSON format. Similarly, any other arbitrary boundaries suiting the sensing application can be handled by Modulo.
    The maps are plotted courtesy of Mapbox.}
    \label{fig:stratification_division_types}
\end{figure*}

%% file: methodology.tex
In this section, we provide a comprehensive description of the datasets used in our evaluation. Further, we present two non-trivial baselines for selecting vehicles from their past mobility data. Next, we will describe the experimental setup in detail. Finally, we introduce the metric used to evaluate the efficacy of Modulo. 

\subsection{Dataset description}
For real-world evaluation of Modulo, we used taxi-tracking datasets for the city of San Francisco~\cite{epfl-mobility-20090224} and Rome~\cite{roma-taxi-20140717} from CRAWDAD. Moreover, we also found public transport transit data for the buses operated by the San Francisco Municipal Transportation Agency~\cite{avl}. These two cities differ from each other in three factors that affect transportation:

\begin{itemize}
    \item Geography -- The two cities are in different continents (San Francisco in N. America and Rome in Europe)
    \item Population density -- Rome: 2,232$/\text{km}^2$ \cite{rome_population}; San Francisco: 7,272$/\text{km}^2$ \cite{sf_population}
    \item City extent -- Rome: 1,285 $\text{km}^2$; San Francisco: 600.59 $\text{km}^2$ \cite{sf_area}
\end{itemize}

The San Francisco taxi dataset contains 11,219,878 records over 25 days in 2008 and reports GPS data every 1 minute. The Rome taxi dataset contained 21,817,850 records over 25 days in 2014, but it reports GPS data every 15 seconds. Hence we consider one in every four records to get minute-level data to maintain consistency across our experiments. Each record in both the taxi-tracking datasets contains a timestamp, taxi ID, and the corresponding GPS location. The San Francisco taxi dataset also contains taxi occupancy information, which we ignore. In both these datasets, we perform our experiments over seven days. The San Franciso bus dataset contained 5,610,179 records over 7 days in 2013. However, for realistic comparison with the San Francisco taxi dataset, we obtain records on the same month and dates as those of the taxi dataset and transpose them from 2013 to 2008.

\begin{table}[h]
    \centering
    \begin{tabular}{|r|c|c|c|}
        \hline
         & \textbf{Records} & \textbf{Vehicles} & \textbf{Time Period}\\\hline
        San Francisco (Bus) & 5603166 & 627 & 7 days\\
        San Francisco (Cabs) & 2977508 & 511 & 7 days\\
        Rome (Cabs) & 631535 & 304 & 7 days\\\hline
    \end{tabular}
    \caption{Summary of datasets used}
    \label{tab:dataset_summary}
\end{table}

\begin{figure*}[t]
    \centering
    \begin{tabular}{cc}
    \includegraphics[height=2in]{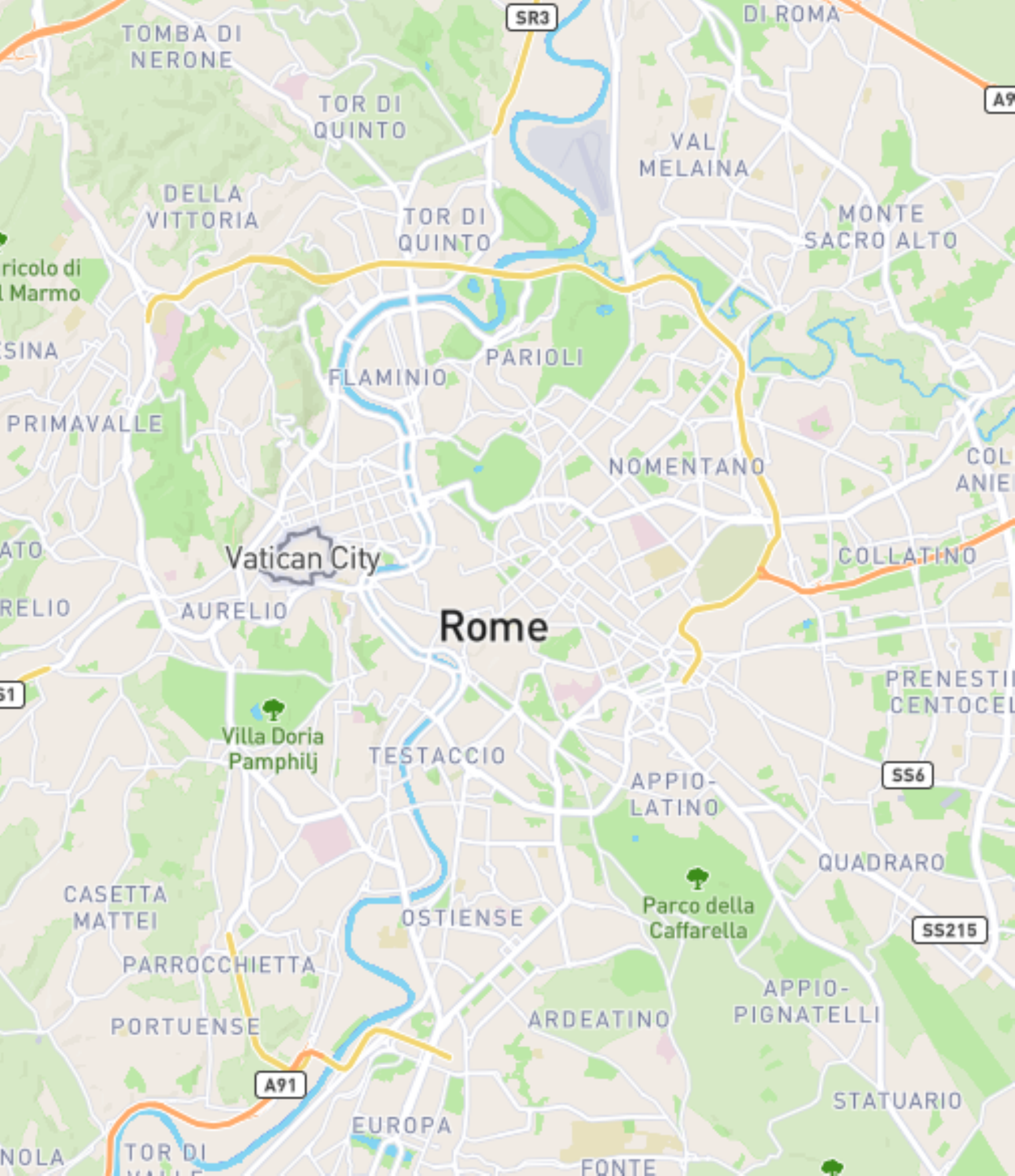} &
    \includegraphics[height=2in]{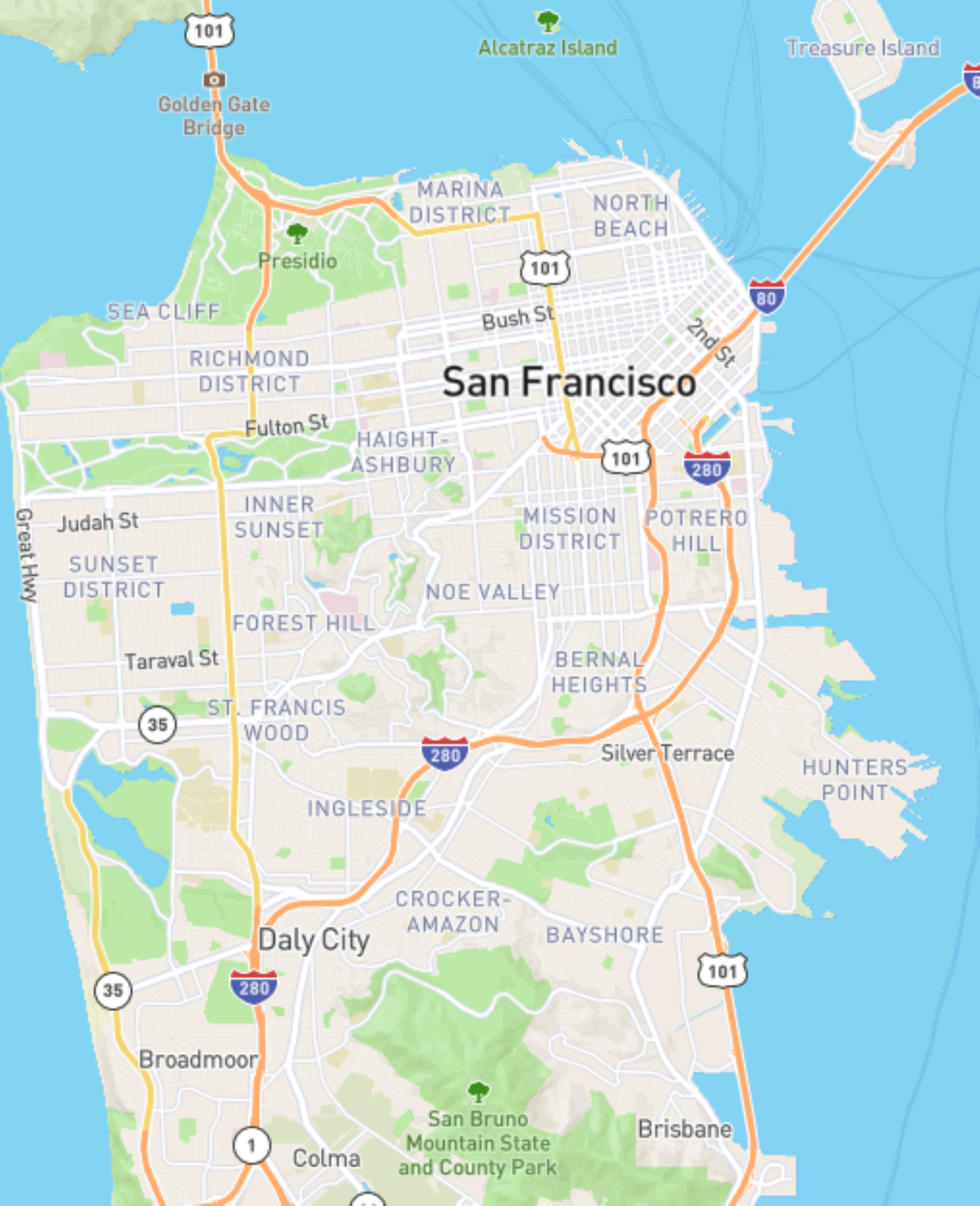} \\
    (a) Selected area of Rome & (b) Selected area of San Francisco\\
    \end{tabular}
    \caption{This figure shows the areas of San Francisco and Rome falling into the selected rectangles. In [minimum longitude, minimum latitude, maximum longitude, minimum longitude] format, the bounds for Rome are: [12.4, 41.825, 12.575, 41.975]. The bounds for San Francisco are: [-122.35, 37.67, -122.515, 37.83]. The maps are plotted courtesy of Mapbox.}
    \label{fig:city_selected_areas}
\end{figure*}

For our experiments, we consider only the records in a rectangle around the center of the city. We do this to limit the number of grids we divide the cities into (as explained in section \ref{sec:implementation}) for computational convenience. The selected rectangle covers about 90\% of the records in both San Francisco and Rome. In terms of area, this rectangle covers 250 $\text{km}^2$ of San Francisco, and 240 $\text{km}^2$ of Rome. In San Francisco, a substantial part of this rectangle covers oceans (for example, the San Francisco Bay) to be able to cover the Golden Gate Bridge and Treasure Island. After selecting the records inside the rectangle over seven days, the dataset that we use for our experiments is summarized in table \ref{tab:dataset_summary}.



In addition to the taxi data, we also consider data from reference-grade air pollution sensors in these two cities. Since we do not know of the existence of the reference-grade sensors in 2008, we use the current (August 2019) locations of these sensors from AQICN.org~\cite{aqicn}. Further, we assume that each of these sensors reports data at a 15-minute interval~\footnote{Reference-grade monitors typically report one value every 15 minutes.}. Here, as well, we only consider the reference-grade sensors located inside our rectangle of interest. According to AQICN.org, there is just one reference-grade monitor in San Francisco and four in Rome. We consider this data in addition to the vehicle data to investigate the number of ``colocation'' or ``rendezvous'' instances of taxis with high-grade air pollution sensors for highly accurate 1-hop calibration \cite{fu2017multihop}.

\subsection{Baseline Description}
In this section, we describe the two baseline methods against which we compare the results of our algorithm.
\begin{itemize}
    \item \textbf{Random - Minimum Points} -- In the Random minus Minimum Points (hereon referred to as Random-MP) selection method,  the required number of cabs are chosen from the set of vehicles reporting $\geq k$ records.
    \item \textbf{Maximum Points} -- In this method, the vehicles reporting the maximum number of records in the vehicle mobility dataset are selected without considering any other heuristic.
\end{itemize}


\subsection{Experimental Setup}
In this section, we explain the setup we use for performing our experiments. While using Modulo, we chose 'grids' as the division type. Further, we input the temporal granularity to be 2 hours and the spatial to be 100 meters. 


For a fair comparison among the three methods (Modulo + 2 baselines), we divide the seven days of the taxi and bus data into two halves of equal periods. This splitting is akin to the training and test set in case of evaluating machine learning algorithms. The data corresponding to the first period was used by the three methods to compute the list of selected vehicles. We evaluated the three methods by using the vehicles selected on the data corresponding to the second period.

\subsection{Evaluation Metric: Percentage Coverage}
We design an intuitive metric to compare the effectiveness of Modulo with the other baselines. Let $N$ be the set of all vehicles present in a dataset. Also, let a techniques select $M$ ($M \subset N$) vehicles. Further, similar to the notations used in section~\ref{sec:algorithm}, we have $C_{i,d,t}$, which is a binary variable representing if the $i^{th}$ vehicle was in the $d^{th}$ segment at time interval $t$. Thus, we define the metric \textbf{Percentage Coverage}, as follows:

\begin{align*}
\text{Percentage Coverage} = 100 \cdot \frac{\sum\limits_{i \in M, d \in D, t \in T} C_{i,d,t}}{\sum\limits_{j \in N, d \in D, t \in T} C_{j,d,t}}
\end{align*}

%% file: results.tex
In evaluation, we seek to gain insights on the following questions - (i) How effective is Modulo in selecting vehicles on different datasets containing different vehicle types?, (ii) How effective is Modulo in running spatio-temporal queries?

\subsection{Baseline Comparisons over different vehicle types}
\begin{figure*}[t]
\centering
\begin{tabular}{cc}
\includegraphics[width=3in]{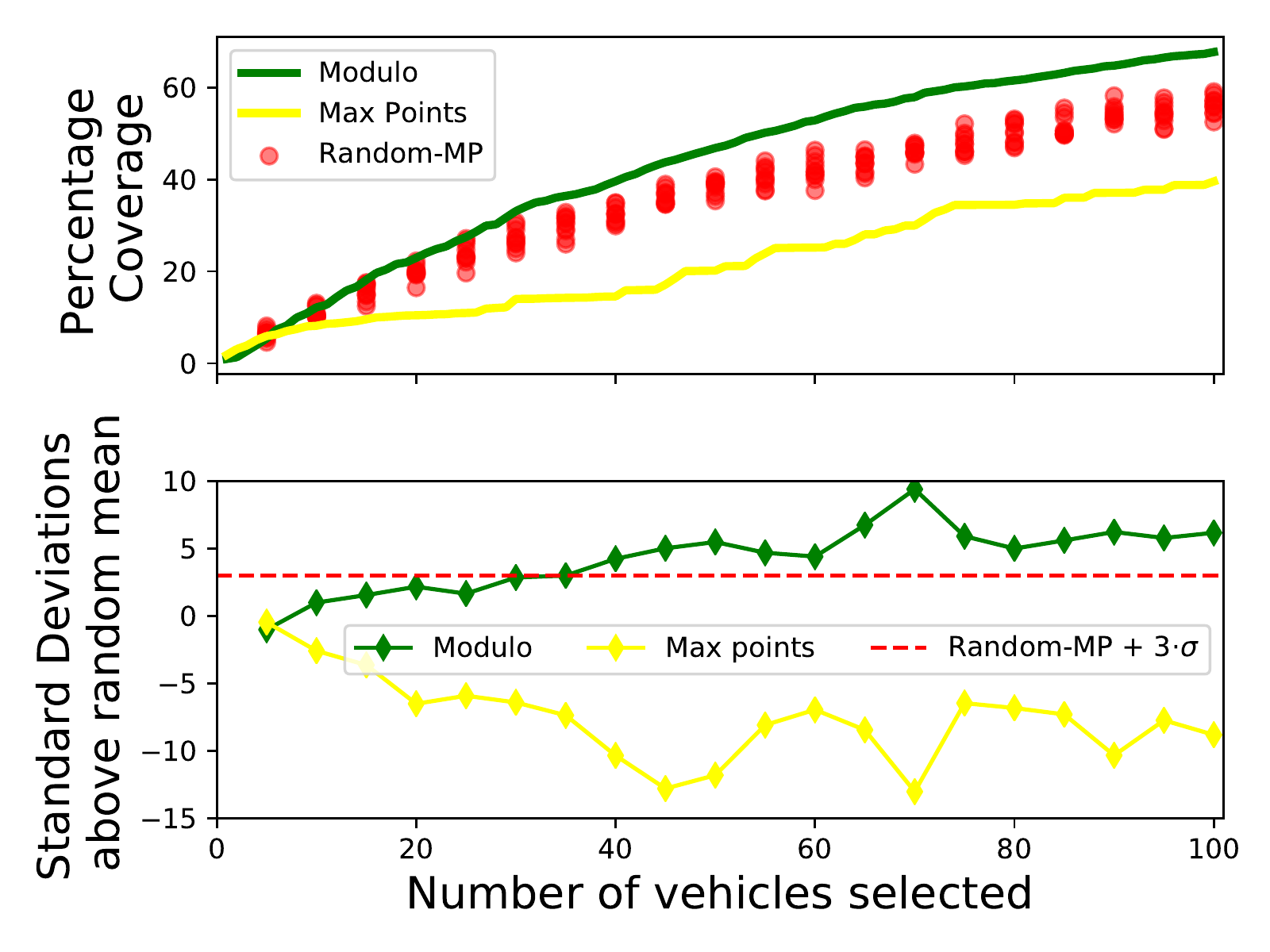} &
\includegraphics[width=3in]{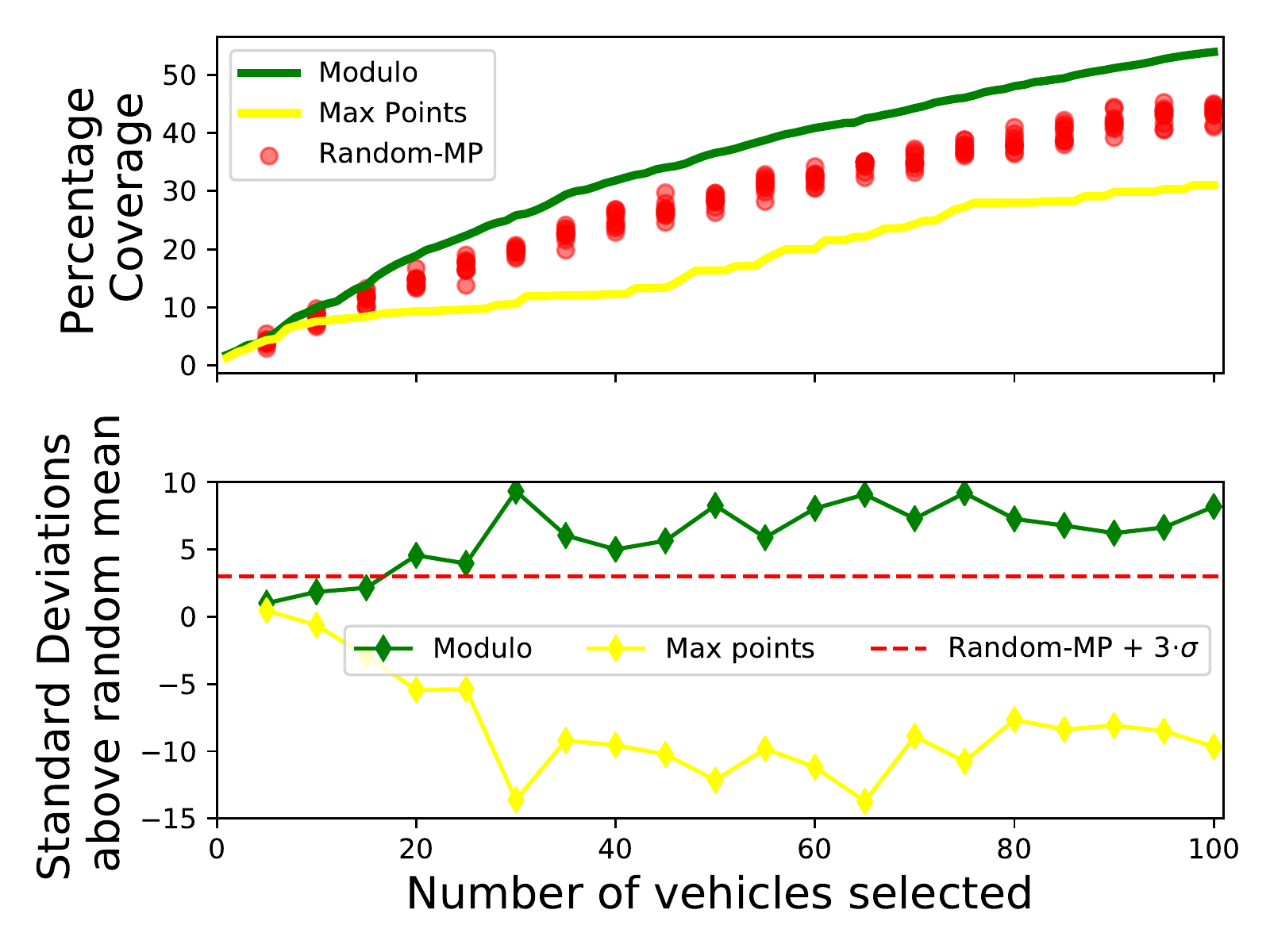} \\
(a) San Francisco (Bus)  & (b) San Francisco (Mix)  \\
\includegraphics[width=3in]{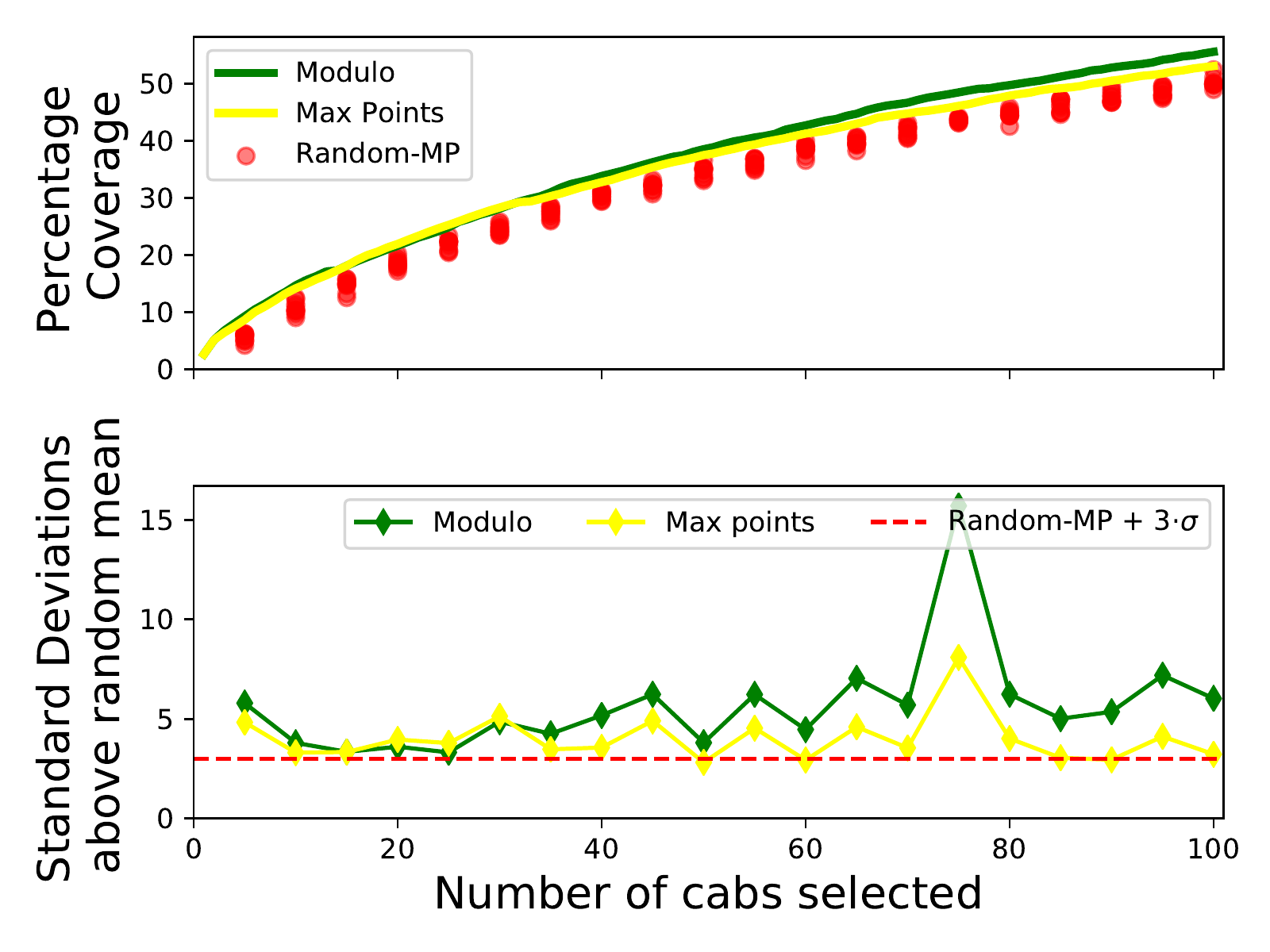} &
\includegraphics[width=3in]{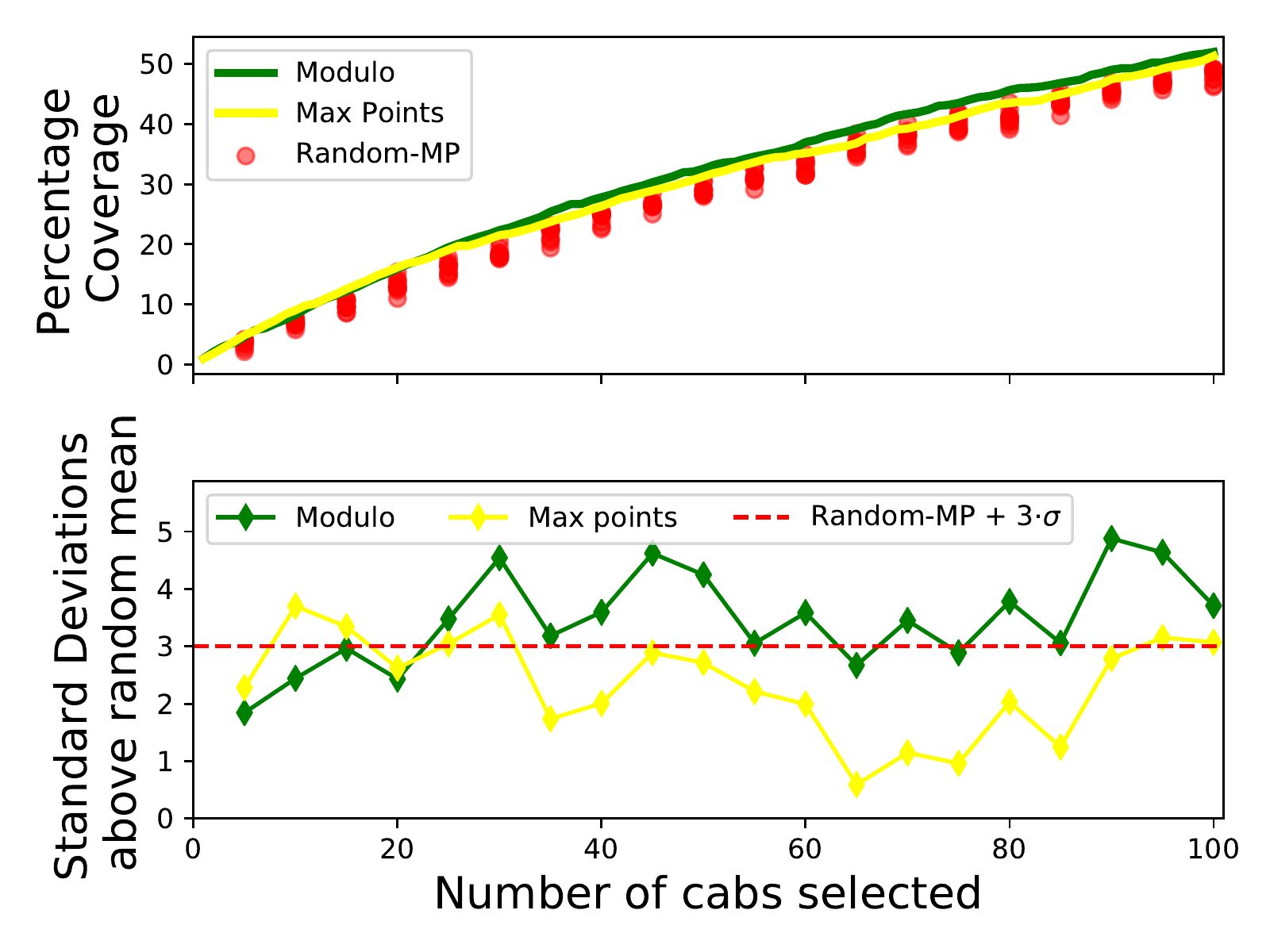} \\
\end{tabular}
\caption{Coverage through drive-by sensing for San Francisco and Rome (Modulo vs Max Points vs Random-MP)}
\label{fig:coverage_results}
\end{figure*}

We examine the performance of our approach, Modulo, as compared to the performance of the two baselines, Random-MP, and Max Points. To obtain these results, we first divide our dataset into two halves of 3.5 days each. We then allow each of the three approaches to select vehicles from the first set that must be deployed in the second set for increased coverage. Finally, we calculate the performance metric of \textbf{percentage coverage} obtained by the selected vehicles from each approach in the second set. We repeat the experiment for a selection budget of up to 100 vehicles. In case of Random-MP, we runs 10 experiments with different seeds for different deployment sizes in the multiple of 5. 
Figure \ref{fig:coverage_results} shows the results of this experiment on 4 datasets: San Francisco bus dataset, San Francisco mixed dataset (cabs and buses together), San Francisco cabs dataset, and Rome cabs dataset.

Figure \ref{fig:coverage_results} (a) shows the difference in the performance of the three approaches for the San Francisco bus dataset. The plot on the top indicates that for fewer buses, all the three approaches provide similar percentage coverage. However, as the deployment size increases, the percentage coverage resulting from Modulo starts to overtake the performance of Random-MP and Max Points. Specifically, we note that to achieve 40\% coverage, Modulo requires 39 buses, Random-MP requires 55 buses ($\approx$ 41\% extra), and Max Points requires a whopping 92 buses ($\approx$ 136\% extra). The plot on the bottom shows the difference in performance between Modulo (green line) and Max Points (yellow line) with the mean of the 10 experiments of Random-MP. As seen, for a deployment size of more than 30 buses, Modulo always performs more than 3 standard deviations (indicated as a dotted red line) better than the mean of the Random-MP deployment. This behavior is explained by the fact that buses ply on fixed routes, which results in the predictability of routes in the future. Since Modulo uses historical trends to select cabs, this predictability allows it to provide increased coverage in the future as well. 

Figure \ref{fig:coverage_results} (b) shows similar results when the three approaches are applied to the dataset containing both fixed-route vehicles (buses) and random-route vehicles (taxis). Modulo performs better than Random-MP and Max Points, and offers more than 3 standard deviations higher coverage that the Random-MP baseline. However, in this case, Random-MP performs much worse than in figure \ref{fig:coverage_results} (a) with respect to the performance of Modulo. This is expected since Random-MP now has a mixed set of vehicles to pick randomly from. Hence, it may pick a few cabs, as opposed to picking all buses with predictable routes with higher coverage in figure \ref{fig:coverage_results} (a). Surprisingly, Max Points picked up just 2 cabs in its selection of 100 vehicles, whereas Modulo picked as many as 37 cabs out of the overall 100. This suggests that although buses have predictable routes, they might be spending a lot of time in similar locations and could have sparse coverage. A common trend observed in both Figure \ref{fig:coverage_results} (a) and Figure \ref{fig:coverage_results} (b) is that there is diminishing return when the deployment size increases. 

\textbf{Observation.} \textit{ Modulo out-performs both Random-MP and Max Points when fixed route vehicles are allowed to be picked because they allow applicability of past trends to the future. However, even in case of mixed routes, Modulo picks vehicles based on their past coverage and thus performs significantly better. As the vehicle count increases, there are diminishing returns in percentage coverage, i.e., the increase in the deployment size follows a sub-linear trend.}

Figure \ref{fig:coverage_results} (c) and (d) show the performance results of the three approaches in datasets comprising only of taxis, i.e., San Francisco and Rome, respectively. In this scenario, there is only 2-5\% increase in the percentage coverage offered by Modulo over Max Points, and 7-11\% increase over Random-MP, when the size of deployment is higher than 40\%. This difference is performance is lower for smaller deployments. 
This result suggests that the historical movement data of taxis is not indicative of the trends they would follow in the future. Neither of the three approaches benefits from any predictability in the movement patterns. In San Francisco, Modulo and Max Points approaches are consistently more than three standard deviations better than the mean values for Random-MP. However, in Rome, Max Points struggles to offer significant improvement, i.e., less than two standard deviations, above the mean.

\textbf{Observation.} Lack of predictability of the movement of taxis results in no significant outperforming of one approach over the other. In these cases, the absolute performance of Modulo over other methods is marginal; there is still a significant statistical difference in performance.

\subsection{Modulo Runtime Results}
In these experiments, we show the efficiency of Modulo in performing spatio-temporal queries on a large database. We perform these experiments on the original databases without removing any records. However, the process of retrieving per-minute values instead of per-15-second values results in the reduction of the Rome dataset from 21,817,850 records to 3,094,358 records. Hence, we perform this experiment on the San Francisco database of size 11,219,878 records and Rome database of size 3,094,358 records.

\begin{figure*}[t]
   \centering
\minipage{0.31\textwidth}
  \includegraphics[width=\linewidth]{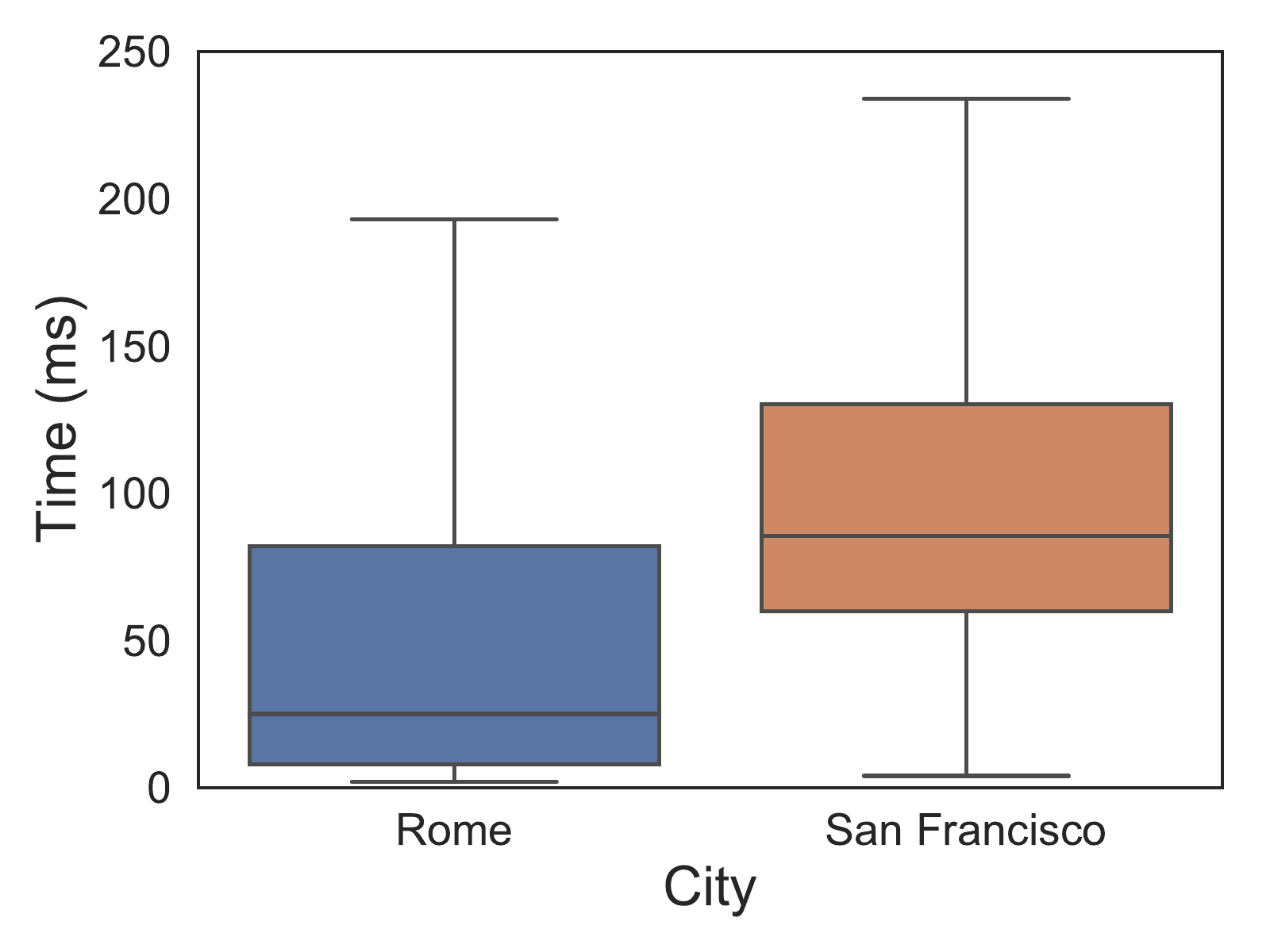}
  \caption{Colocation query time}
 \label{fig:query-time}
\endminipage \hfill
\minipage{0.31\textwidth}
  \includegraphics[width=\linewidth]{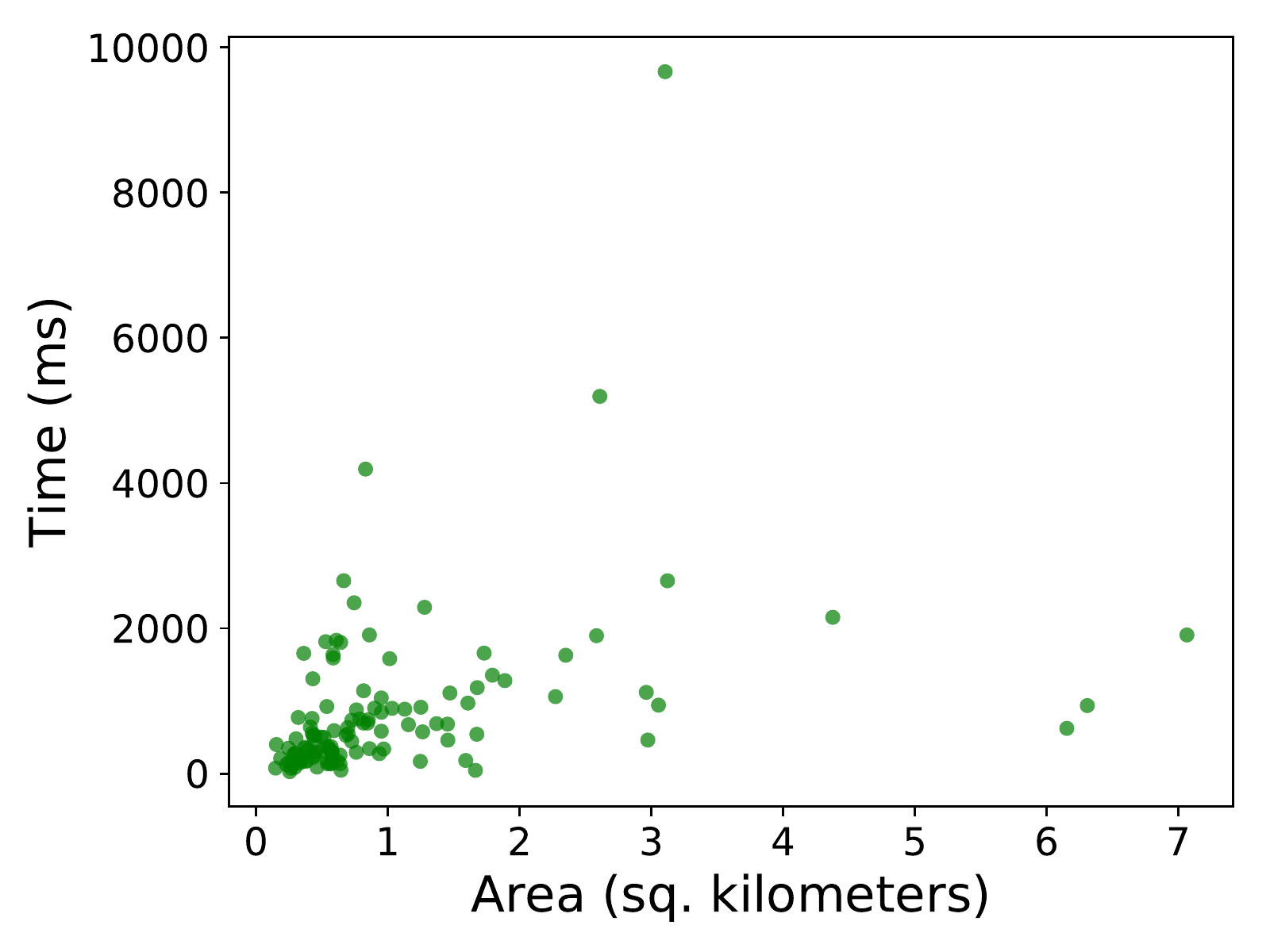}
  \caption{Query time vs Area}
   \label{fig:qt-area}
\endminipage\hfill
\minipage{0.31\textwidth}
  \includegraphics[width=\linewidth]{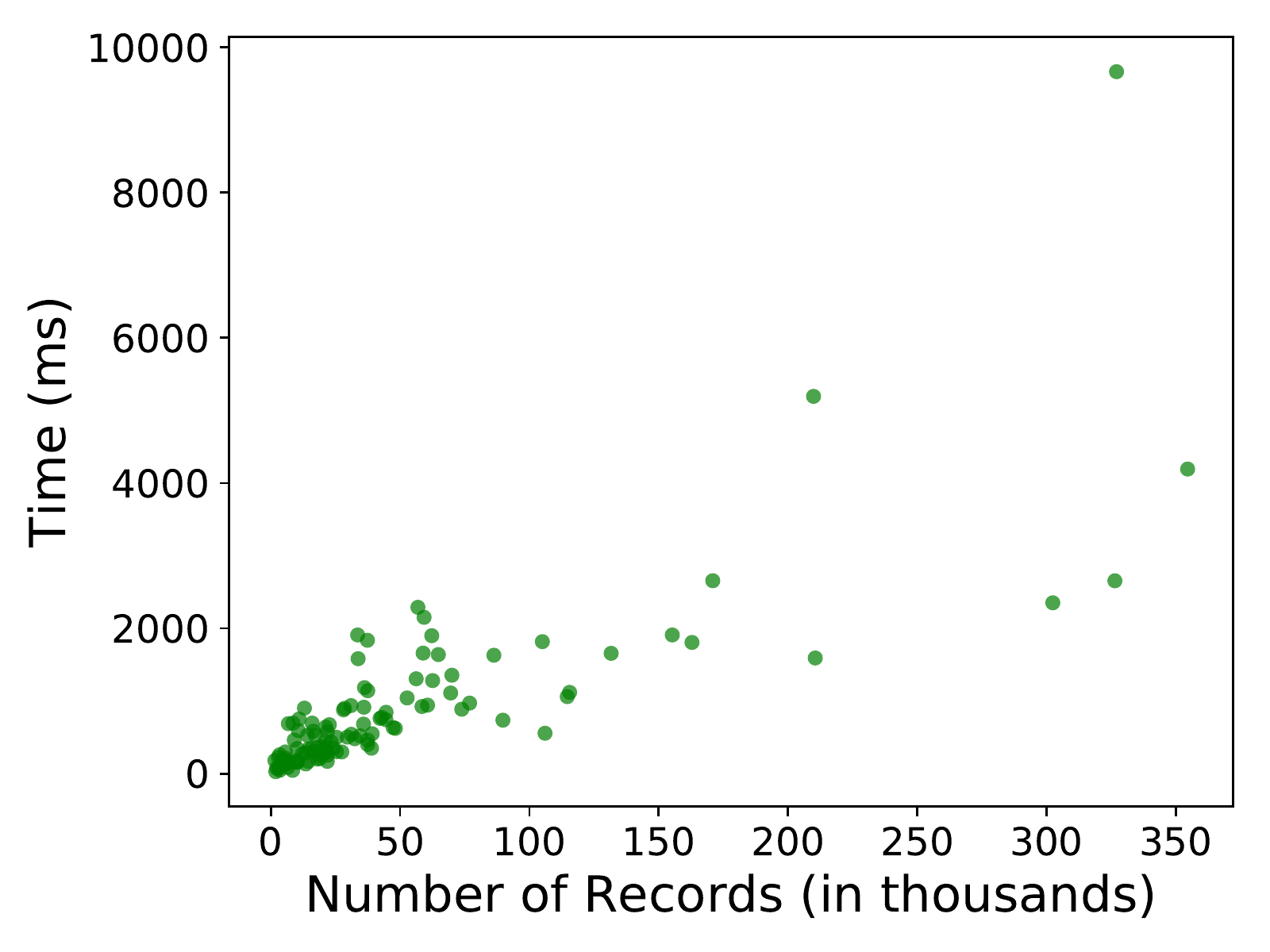}
  \caption{Query time vs Points found}
   \label{fig:qt-points}
\endminipage\hfill
\end{figure*}

For the experiment shown in figure \ref{fig:query-time}, we choose 500 random coordinates associated with records in the San Francisco and Rome databases. For each chosen coordinate, we run a spatio-temporal query to find all colocations around that point, i.e., all points in its 50 meters spatial radius and 5 minute temporal radius. The reported time is the total time required by the database for query plan selection and query execution in milliseconds \cite{mongodb}. The mean query time for the Rome database is 75.55 milliseconds and for the San Francisco database in 99.7 milliseconds. This difference is expected as the San Francisco database is almost 4$\times$ as large as the Rome database.

\textbf{Observation.} \textit{The average query time in the Rome database is 75.55 milliseconds and in the San Francisco database is 99.7 milliseconds. There seems to be a correlation between query time and the size of the database.}

Figure \ref{fig:qt-area} and figure \ref{fig:qt-points} show another experiment on Modulo's geospatial database. In this experiment, we test the performance of the database on arbitrarily sized polygons. Efficient resolving of these queries is important for Modulo to be able to provide in-built support for custom stratification. Further, this allows data stored in Modulo database to be ideally suited for social science experiments involving interventions created in different portions of a given city. To evaluate the performance of Modulo for arbitrarily sized regions, we use a GeoJSON file of polygons defining the neighbourhood boundaries of San Francisco \cite{sf_geojson}. Then, for each neighborhood, we query our database for all the records lying in it. We report the time taken for each such query with respect to two metrics: the area of the neighborhood in figure \ref{fig:qt-area}, and the number of points returned in the neighborhood \ref{fig:qt-points}. The Pearson's $r$ coefficient for the area-time plot is 0.363 and for the number-time plot is 0.797.

\textbf{Observation.} \textit{There seems to be a weak correlation between the area of an arbitrary polygon and the query time. However, there appears to be a stronger correlation between the number of points in a polygon and the query time.}

\subsection{Real-world Deployment using Modulo}
\begin{figure*}[t]
    \centering
    \includegraphics[width=3in]{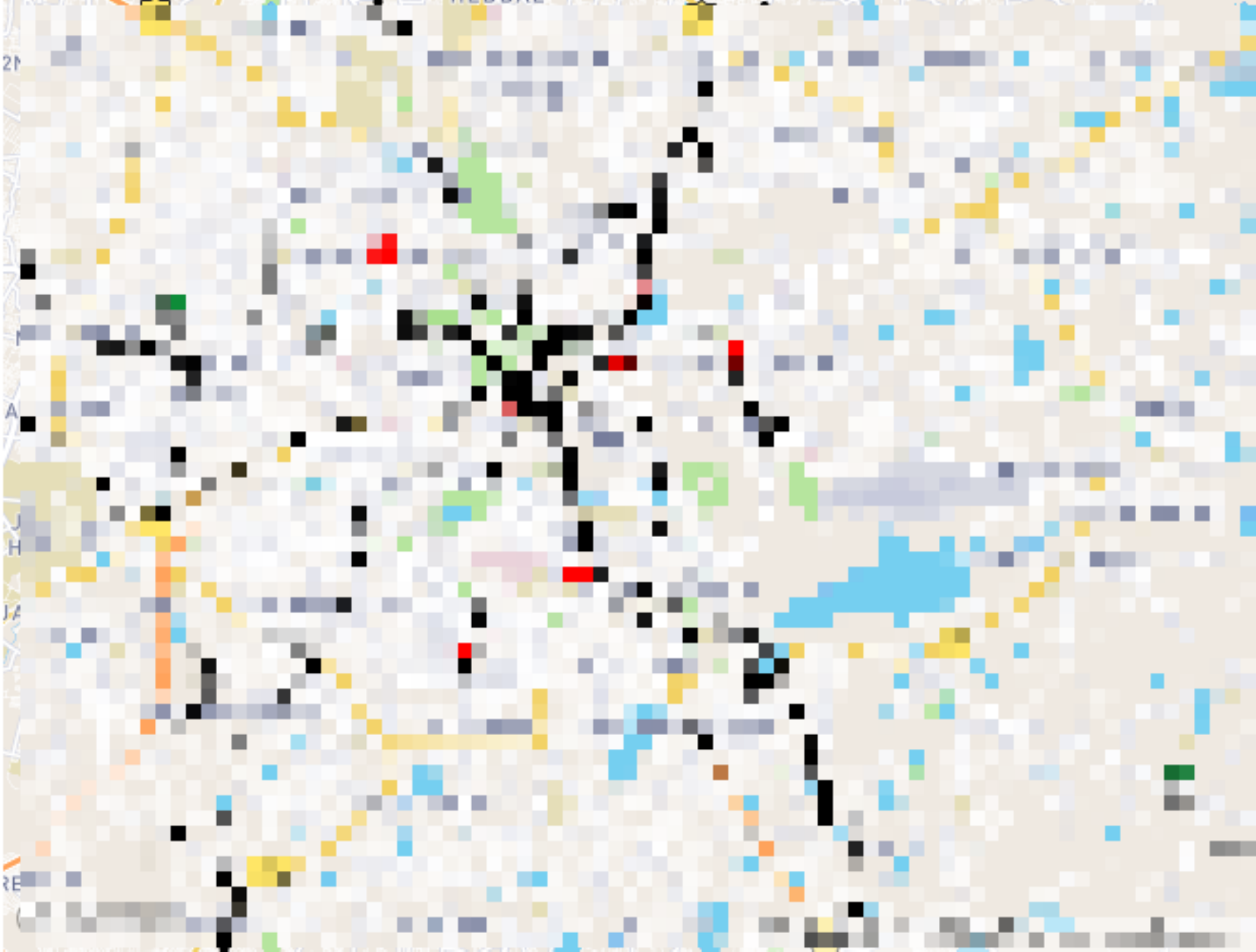}
    \caption{Map showing the records collected in a 12 hour period in our real-world deployment. The black markers represent the collected records. <Image pixelated for anonymity>}
    \label{fig:bangalore_deployment}
\end{figure*}

Here, we present a real-world case-study, where we use Modulo to deploy particulate matter 2.5 (PM2.5) monitors on a few cabs in a Southern Indian city. We had access to the mobility patters of 25 cabs belonging to an organization <removed for anonymity>. These cabs ran for $\approx$2 hours in the morning during the employee check-in time, and $\approx$2 hours in the evening after the employee check-out time. We chose a custom stratification model representing the administrative wards in the city. We used a temporal resolution of 15 minutes as PM2.5 values are very dynamic. We fed all this information into Modulo, which selected 14 of the 25 cabs for maximized coverage. We deployed this network in April 2019 and have collected over 2 million data points till August 2019. A snapshot of our deployment can be seen in the figure \ref{fig:bangalore_deployment}.

%% file: related.tex
\subsection{Urban drive-by sensor deployments}
Several papers have looked at \textit{drive-by sensing} as an effective method for collecting data for various use-cases~\cite{mohan2008nericell,eriksson2008pothole,HSWT2011a,gao2016mosaic,datta2014towards,he2018detecting,rong2018parking,mathur2010parknet}. The applications provide a promising overview of how \textit{drive-by sensing} can be used in different urban environments. Although these are exciting applications, a detailed study on the city-scale viability of their approach using \textit{drive-by sensing} is not discussed. 

Anjomshoaa et al. provided an empirical evaluation on the street coverage obtained from fixed-route and random-route vehicles. However, beyond a few statistical results, this work does not present a complete strategy on selecting an optimal set of vehicles based on their study. The work presented by Ali et al.~\cite{ali2017coverage} looked at the application of pothole detection and modeled the sensor deployment as a maximum coverage problem. However, they only looked at vehicles having pre-defined routes and do not consider the various extensions needed for serving several use-cases for all kinds of \textit{drive-by sensing} application. 

Liu et al. presented a detailed survey on a broader topic of mobile crowdsensing~\cite{liu2018survey}. They also talk about several works using \textit{drive-by sensing}. But the sensor selection problem for \textit{drive-by sensing} is not discussed. A recent work by O'Keefe et al.~\cite{o2019quantifying} show interesting statistical results on cabs registered on ride-sharing and city taxis. However, they do not go beyond the random selection of taxis in their evaluation.

Another recent work explores the idea of sensor reliability at a large scale using a sensor voltage-based primitive called the ``Fall-curve''~\cite{chakraborty2018fall}. This approach is complementary to the work presented in this paper, and there can be possible synergies with similar approaches that can work at a city-scale. 

\subsection{Air Quality sensing}
An exciting use-case of \textit{drive-by sensing} is for air quality monitoring. This application has several unique challenges, such as on-field calibration of the gas and PM sensors, localized phenomena of pollution, etc.  Several papers have utilized sensors deployed on moving vehicles to collect pollution data~\cite{fu2017multihop,gao2016mosaic}. Air quality monitoring is a perfect application for \textit{drive-by sensing} as pollution is a local phenomenon, and there are significant variations from one place in the city to the other. Interestingly, Fu et al. have looked at the optimal placement of expensive monitors to make low-cost sensors k-hop calibrable~\cite{fu2017multihop}. They formulate the problem as a set cover problem. However, these results on k-hop calibrability apply only to fixed routes vehicles. Saukh et al. also looked at the route selection problem to achieve coverage for fixed-route vehicles~\cite{SHNUT2012a}.

%% file: discussion.tex
According to the UN's Department of Economic and Social Affairs report, our cities will house greater than 68\% of our population by 2050, compared to today's 55\%~\cite{undev}. Such massive influx is primarily going to be severe in Asian and African cities. Economist Intelligence Unit’s (EIU) annual survey (``The Global Liveability Index") consistently ranks these cities at the bottom of their list~\cite{eiu}. Especially in these cities, \textit{drive-by sensing} holds significant potential as it enables sensing a geographically wide region using just a few hundred sensors. Mobile sensing can provide probes to gauge the health of a city while dramatically reducing the cost of deployment. Further, they would unleash a plethora of analytics that can be performed to improve the livability of our cities. Below, I provide some exciting avenues to explore the ideas presented in this paper. 

\subsection{Multi-modal sensing}
Each vehicle available for \textit{drive-by sensing} can carry several sensors at once. Past work has looked at using an array of sensors for measuring individual pollutants present in our cities~\cite{maag2017scan}. Such \textit{sensor overloading} can further drive down the cost of deployment and enable compelling opportunities for sensor fusion. An interesting extension to our current work could be to select vehicles for multi-modal sensing with differing spatio-temporal granularities.

\subsection{Real-time sensing and calibration}
As shown, \textit{drive-by sensing} yields decent coverage with a fraction of vehicles. To ensure better coverage, we can send a small set of hired vehicles to segments of the cities not visited by our original fleet. Moreover, equipping such vehicles with recently calibrated sensors will ensure better calibration of static and other mobile sensors along the path. Operating hired vehicles is akin to the \textbf{vehicle routing problem}, a well-known problem in theoretical computer science. 

\subsection{Use of Multi-Objective or Robust optimization}
Our integer programming formulation looked at optimizing a single objective, i.e., either coverage maximization or budget minimization. However, there could be cases where we would like to optimize two or more objectives simultaneously. Consider a scenario, we might want to minimize the number of vehicles needed while simultaneously reducing disparity in sensing economically-backward parts of a city. Multi-objective programming is ideally suited to tackle such instances. Further, one can model soft constraints. For example, many cities in the developing world have few, if any, reference-grade pollution monitors. Robust optimization allows some constraints --- such as colocations with reference monitors --- to be unmet for a small fraction of selected vehicles.



%% file: conclusion.tex
\textit{Drive-by sensing} is an upcoming way of sensing the physical phenomena around us with fine-granularity at a city. But for city-scale deployments to be successful, the vehicle fleet used for the deployment needs to be carefully selected so that it fulfils all the requirements of the sensing application like coverage at a specified spatial and temporal granularity, colocations for calibration, dynamism in the deployment, etc. In this paper, we proposed Modulo -- a novel approach to the deployment of vehicles for \textbf{drive-by sensing} that is generalize to multiple sensing applications. We expose a Python library that enables sensor network operators to use this approach for selecting the optimal set of vehicles from a fleet of candidate vehicles. We also propose variations of our algorithm to suit specific needs like budgeted deployment and weighted coverage. We compare our approach against a couple of baseline algorithms on three kinds of datasets: all cabs, all buses, a mix of cabs and buses. We see that our algorithm outperforms the other baseline methods. Specifically, for a \textit{drive-by sensing} application in the city of San Francisco, we obtained 40\% coverage using just 39 public transport buses. We also conducted a real-world case-study where we used Modulo to select vehicles for an air pollution sensing application. 